\documentclass{aa}  
\usepackage{geometry}
\usepackage{graphicx}
\usepackage{placeins}
\usepackage{txfonts}
\usepackage[table,svgnames,dvipsnames]{xcolor}
\usepackage{hyperref}
\definecolor{darkraspberry}{rgb}{0.53, 0.15, 0.34}

\def\der{\mathrm{d}}

\begin{document}

   \title{Impact of assembly bias on clustering plus weak lensing cosmological analysis}


   \author{R. Paviot 
          \inst{1}\thanks{e-mail: romainpaviot@gmail.com}
          \and
          A. Rocher\inst{2,3}
          \and
          S. Codis \inst{1}
          \and
          A. de Mattia \inst{2}
          \and
          E. Jullo \inst{4}
          \and
          S. de la Torre \inst{4}
          }

   \institute{Université Paris-Saclay, Université Paris Cité, CEA, CNRS, AIM, 91191, Gif-sur-Yvette, France
         \and
Université Paris-Saclay, CEA, Département de Physique des Particules, 91191, Gif-sur-Yvette, France
\and
Laboratory of Astrophysics, École Polytechnique Fédérale de Lausanne (EPFL), Observatoire de Sauverny, CH-1290 Versoix, Switzerland 
\and
Aix Marseille Univ, CNRS, CNES, LAM, Marseille, France \\
             }

   \date{Received XXX; accepted XXX}

 
  \abstract
  {Empirical models of galaxy-halo connection such as the halo occupation distribution (HOD) model have been widely used over the past decades to intensively test perturbative models on quasi-linear scales. However, these models fail to reproduce the galaxy-galaxy lensing signal on non-linear scales, over-predicting the observed signal by up to $40 \%$.}
   {With ongoing Stage-IV galaxy surveys such as DESI and \textit{Euclid} that will measure cosmological parameters at sub-percent precision, it is now crucial to precisely model the galaxy-halo connection in order to accurately estimate the theoretical uncertainties of perturbative models.}
   {This paper compares a standard HOD (based on halo mass only) to an extended HOD that incorporates as additional features galaxy assembly bias and local environmental dependencies on halo occupation. These models were calibrated against the observed clustering and galaxy-galaxy lensing signal of eBOSS luminous red galaxies and emission line galaxies in the range $0.6 < z < 1.1$. We performed a combined clustering-lensing cosmological analysis on the simulated galaxy samples of both HODs to quantify the systematic budget of perturbative models.}
   {By considering not only the mass of the dark matter halos but also these secondary properties, the extended HOD offers a more comprehensive understanding of the connection between galaxies and their surroundings. In particular, we found that the luminous red galaxies preferentially occupy denser and more anisotropic environments. Our results highlight the importance of considering environmental factors in empirical models with an extended HOD that reproduces the observed signal within $20 \%$ on scales below 10 $h^{-1}$Mpc. Our cosmological analysis reveals that our perturbative model yields similar constraints regardless of the galaxy population, with a better goodness of fit for the extended HOD. These results suggest that the extended HOD should be used to quantify modelling systematics. This extended framework should also prove useful for forward modelling techniques.}
   {}

   \keywords{Galaxies: halos --  Galaxies: statistics
                (Cosmology:) large-scale structure of Universe
               }

\titlerunning{Impact of assembly bias on cosmological analysis}

   \maketitle
%

\section{Introduction}
In the standard model of cosmology, galaxies are predicted to form and evolve within dark matter halos. To extract cosmological information from observed galaxy clustering, it is necessary to understand galaxy formation and thus the connection between galaxies and their host dark matter halos beyond their linear relationship on large scales \citep{1984ApJ...284L...9K}. Analytically, one can go beyond the linear approximation to describe the relationship between matter and galaxies \citep{Mcdo2009,assassi14}. Unfortunately, these perturbative models fail to describe the connection between matter and galaxy at intra-cluster scales, thus requiring the use of empirical models calibrated on simulations. This is even more important nowadays, as forward modelling techniques have recently emerged as a means to constrain cosmology \cite{Huan2023,kidssbi2024,dessbi2024}.

The two most efficient and popular models of galaxy-halo connection for large-scale cosmological studies are the sub-halo abundance matching (SHAM) model \citep{Conroy2006,Vale2006,Reddick2013} and the halo occupation distribution \citep[HOD;][]{Berlind2002,Zheng2005,Zheng2007} model. SHAM assumes a direct correspondence between sub-halo mass and the stellar mass or luminosity of its hosted galaxies and can reproduce the clustering of stellar mass-selected samples in hydrodynamical simulations with high accuracy \citep{Chaves2016}. Similarly, HOD models can reproduce the clustering of magnitude-selected samples and have been commonly used to produce realistic galaxy catalogues for the Sloan Digital Sky Survey \citep[SDSS;][]{York2000} observations \citep{White2011,Zhai2017,Alam2020,Avila2020,Rossi2021}.
In the standard HOD framework, it is assumed that all galaxies live inside dark matter halos and that galaxy occupation, which corresponds to the probability of a halo to host a galaxy, is solely determined by halo mass. This assumption has emerged as halo mass is the halo property that most strongly correlates with halo abundances and clustering as well as galaxy intrinsic properties \citep{White1978}. However, a large number of studies have shown that the mass-only HOD does not properly model anisotropic galaxy clustering on small scales \citep{Yuan2021,Yuan2022} and that halo occupation depends on secondary halo properties beyond halo mass \citep{Artale2018,Zehavi2018,Bose2019,Hadzhiyska2020,Hadzhiyska2023b,Hadzhiyska2023a}. 

In particular, it has been shown that halo occupation and clustering are correlated with properties other than halo mass, such as halo concentration, formation time, and spin -- a phenomenon referred to as halo assembly bias \citep{Gao2007,2008ApJ...687...12D}. This physical process also introduces a dependence of galaxy clustering on properties other than halo mass \citep{Croton2007}, the so-called galaxy assembly bias, which also depends on halo occupation and galaxy formation models \citep{Zehavi2018}. Additionally, galaxy occupation also depends on environmental properties, such as local density and local anisotropies, as one can predict from first principles, for instance, when using excursion set theory \citep{Musso2018}. The previous study of \cite{Yuan2021,Yuan2022} shows that the standard framework fails to accurately model the full two-dimensional redshift space clustering of the  SDSS BOSS CMASS (constant mass) sample \citep{Dawson2013} and that an extended HOD framework that incorporates assembly bias and environmental dependence of halo occupation agrees better with the data. Furthermore, \citep{Leauthaud2017} observed a discrepancy of $30-40 \%$ between the observed galaxy-galaxy lensing (GGL) signal of CMASS galaxies and the standard HOD framework, a tension referred to as 'lensing-is-low'. This tension can be resolved with an extended SHAM model, as demonstrated in \citep{Contreras2023}, so as to account for the impact of galaxy physics \citep{2019A&A...626A..72G,2023MNRAS.521..937C}. In this analysis, we investigate if an extended HOD framework can provide similar improvement to the observed clustering and GGL signal of lens galaxies of the extended BOSS survey \citep{Dawnon2016} around the Dark Energy Survey (DES) Y3 sources \citep{DES2016}. Additionally, we aim to investigate if small-scale variations between different HOD frameworks provide notable changes in the extraction of cosmological information by fitting the clustering and lensing signal with perturbative models.  

The paper is organised as follows: In Sect. \ref{sec:data}, we briefly present the observations. In Sect. \ref{sec:hod}, we describe the implementation of the standard and extended HOD framework as well as our fitting procedure. Finally, we present the results of the HOD and cosmological fits in Sect. \ref{sec:hodfit} and \ref{sec:cosmofit}, respectively. We conclude in Sect. ~\ref{sec:conclusion}, and we also provide additional information in Appendices \ref{sec:appendixA},  \ref{sec:appendixB}, and \ref{sec:appendixC}.  
\begin{figure*}
    \centering    
    \includegraphics[width=17cm]{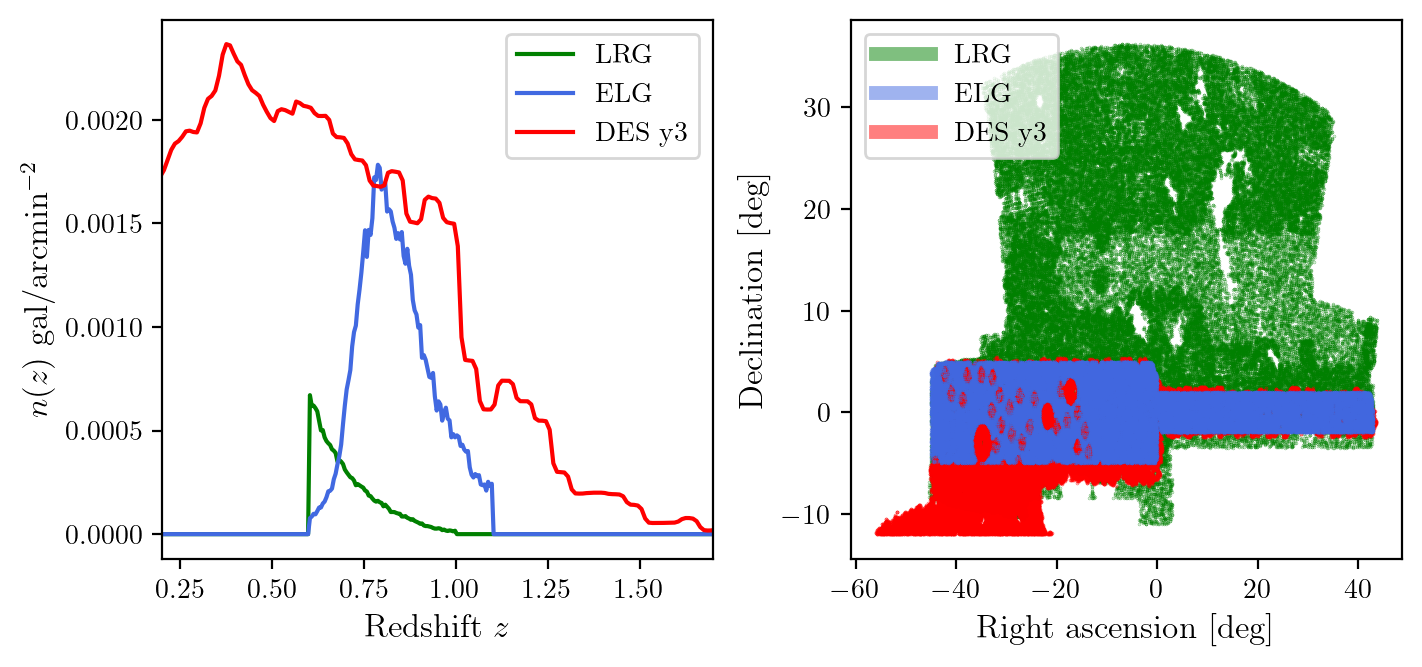}
    \caption{Galaxy number density (left panel) and footprint (right panel) of the LRG sample (green), the ELG sample (blue), and the DES y3 GOLD catalogue (red). The $n(z)$ of DES corresponds to the concatenation of the $n(z)$ on each tomographic bin \citep{Myles2021}. It has been sub-sampled by a factor 20 for clarity. On the right panel, we present the DES footprint with a cut below a declination of $-12.$
      }
    \label{fig:survey_plot}  
\end{figure*}

\section{Data} \label{sec:data}
\subsection{eBOSS and DES catalogues}
The extended Baryon Oscillation Spectroscopic Survey \citep[eBOSS;][]{Dawnon2016} measured thousands of spectra of galaxies and quasars in the range $0.6 < z < 2.4$ in order to probe the geometry and growth rate of the Universe over cosmic time. In this work, we study the properties of the emission line galaxies (ELGs) and eBOSS luminous red galaxies (LRGs) combined with the BOSS CMASS sample.\footnote{We refer to it as an LRG sample for simplicity. Catalogues can be found at \url{https://data.sdss.org/sas/dr17/eboss/lss/catalogs/DR16/}.} Description of these samples can be found in \cite{ross20} and their cosmological analyses in \cite{tamone20,demattia21,bautista21,Hector2021}. We focused our attention on the South Galactic Cap (SGC), where 121 717 LRGs and 89 967 ELGs were observed in the range $0.6 < z < 1.0$ and $0.6 < z  < 1.1$.  In addition to the clustering properties of these galaxies, we aim to extract information from the GGL signal. This is achieved with the shape measurements of DES Year 3 (Y3). We used the main Y3 DES catalogue (referred as 'Gold')\footnote{\url{https://des.ncsa.illinois.edu/releases/y3a2}} --  information on the photometry and shape measurements of this sample can be found in \cite{Sevilla2021,Gatti2021}, while the corresponding cosmological 3x2pt analysis is presented in \cite{DES2022}. The radial distribution and the footprint of each survey are presented in Fig \ref{fig:survey_plot}. eBOSS and DES overlap in a small portion, around 754 deg$^2$, in the SGC. In the overlapping area, 14 413 244 sources can be used to infer the GGL signal of 44 456 LRGs lenses, while all of the observed ELGs are within the DES footprint.\footnote{These values were computed with Healpix \citep{Gorski2005} with NSIDE = 512} Although there are more eBOSS galaxies in the North Galactic Cap, there is no overlapping weak lensing survey with precise and robust shape measurements to measure the GGL signal in this pole. Therefore, we only used the clustering measurement in the SGC so that the clustering and lensing signals are correlated.

\subsection{The UCHUU simulation} \label{sec:uchuu}
In order to model the observed signals, we relied on the N-Body simulation {\sc uchuu} \citep{Ishiyama2021}. This simulation was run with GreeM, a massively parallel TreePM code \citep{Ishiyama2009,Ishiyama2012}, with initial conditions produced by second-order Lagrangian perturbation theory starting at a redshift $z=127$. The cosmological flat $\Lambda$CDM parameters of the simulation are compatible with the ones measured by ~\cite{Planck2015}, ($\Omega_{\mathrm{m}},\Omega_{\mathrm{b}}, h, n_s, \sigma_8) = (0.3089,0.0486, 0.6774, 0.9667,0.8159)$. We note that {\sc uchuu} is one of the most precise N-Body simulations to date, with 12 800$^{3}$ dark matter particles with mass $3.27 \times 10^8 h^{-1}$ $M_{\odot}$ in a box of side length $L=2 h^{-1}$Gpc. This large volume enabled us to study a large number of dark matter halos up to a very high mass of  $M \approx10^{14.8} h^{-1}$ $M_{\odot}$. These halos were identified with the RockStar halo finder \citep{Behroozi2013}. We extracted from the simulation\footnote{\url{https://skiesanduniverses.iaa.es/Simulations/Uchuu/}} the halo and sub-halo catalogues of mass $M >10^{11} h^{-1}$ M$_{\odot}$ and the particle distribution (sub-sampled by a factor 0.005) at redshifts $z=0.86$ and $z=0.7$, corresponding to the effective redshift of the ELG and LRG samples, respectively. 
\section{Halo occupation model}  \label{sec:hod}
\subsection{Standard halo occupation distribution parametrisation}
\label{sec:standardHOD}
In order to simulate realistic galaxy distributions in the {\sc uchuu} simulation, we populated halos using the standard HOD description. 
For LRGs, we considered the HOD framework first described in \cite{Zheng2005,Zheng2007} and widely used in previous BOSS and eBOSS studies \citep{White2011,Zhai2017}. For emission line galaxies, we used the Gaussian HOD model (GHOD) \citep{Avila2020,Rocher2023a} for central occupation. Hence, we defined the mean central halo occupations as\footnote{Throughout this paper, the logarithm notation refers to the decimal logarithm unless specified otherwise.}
\begin{align}
\left\langle N_{\mathrm{cen }}^{\text{LRG}}(M)\right\rangle&=\frac{f_{\text{ic}}}{2}\left[1+\operatorname{erf}\left(\frac{\log M-\log M_{\min }^{\text{LRG}}}{\sigma_{\mathrm{M},{\text{LRG}}}}\right)\right], \\
\left\langle N_{\mathrm{cen }}^{\text{ELG}}(M)\right\rangle&=\frac{A_c}{\sqrt{2 \pi} \sigma_{\mathrm{M},\text{ELG}}} \cdot \exp{-\frac{\left(\log M -\log M_{\min }^{\text{ELG}}\right)^2}{2 \sigma_{\mathrm{M},\text{ELG}}^2}},
\end{align}
where $f_{\text{ic}}$ and $A_c$ correspond to completeness factors and the $\sigma_{\mathrm{M}}$ correspond to scatters of central occupations. When the samples are complete, $f_{\text{ic}}$=1, $M_{\min }^{\text{LRG}}$ is the halo mass at which half of the halos have a central galaxy. $ M_{\min }^{\text{ELG}}$ always corresponds to the halo mass with maximal probability to host a central galaxy. The satellite occupation is defined as 
\begin{align}
\left\langle N_{\mathrm{sat}}(M) \right\rangle= f(M)\left[\frac{M-\kappa M^{\text{LRG}/\text{ELG}}_{\mathrm{min}}}{M_1^{\text{LRG}/\text{ELG}}}\right]^\alpha, \\
\text{with} \  f(M) = \left\{
    \begin{array}{ll}
        \left\langle N^{\mathrm{LRG}}_{\mathrm{cen}}(M) \right\rangle \ \text{for LRG}, \\
         A_s = \mathrm{cst} \ \text{for ELG},
    \end{array}
\right.
\end{align}
where the $M_1$\footnote{We drop the subscripts for simplicity in the following sections.} are the halo masses for which a halo has one satellite galaxy (when the samples are complete), $\alpha$ is the power-law index, and $\kappa$ is a parameter that allows the cut-off of satellite occupation to vary with halo mass.
For the ELG HOD, we assumed the relations given in Table 2 of  \cite{Avila2020},
\begin{equation}
\kappa = 0.8 \ \text{and} \log M_1 =  \log M_{\min } + 0.3.
\end{equation}
From these mean occupations, the number of central galaxies per halo of mass $M$ then
follows a Bernoulli distribution with a mean probability equal to $\left \langle N_{\mathrm{cen }} \right \rangle$. These galaxies are positioned at the centre of their host halos. The number of satellite galaxies per halo follows a Poisson distribution with mean $\left\langle N_{\mathrm{sat}}\right\rangle$, with positions sampled from a Navarro-Frank-White (NFW) profile \citep{NFW96}. Satellite velocities are normally distributed around the mean halo velocity, with a dispersion being equal to the ones of the dark matter halo particles,
\begin{equation} \label{eq:velocitysat}
\vec{v}_{\mathrm{sat}} \sim \mathcal{N}\left(\vec{v}_h, \cdot \sigma_{v_h}\right).
\end{equation}

Eventually, the final number density of mock galaxies can be computed as 
\begin{equation}\label{eq:ngal}
\bar{n}_{\mathrm{gal }}=\int \frac{\mathrm{d} n(M)}{\mathrm{d} M}\left[\left\langle N_{\mathrm{cent }}(M)\right\rangle+\left\langle N_{\mathrm{sat }}(M)\right\rangle\right] \mathrm{d} M,
\end{equation}
and the satellite fraction is given by 
\begin{equation} \label{eq:fsat}
f_{\text {sat }}=\frac{1}{\bar{n}_{\text {gal }}} \int \frac{\mathrm{d} n(M)}{\mathrm{d} M}\left\langle N_{\text {sat }}(M)\right\rangle \mathrm{d} M \text {. }
\end{equation}
Obviously, the mean number density of galaxies is fixed, which gives a constraint on the free parameters of the HOD.
For ELG, we followed the methodology described in \cite{Rocher2023a}. As the mock clustering is governed by the ratio $A_s/A_c$, $A_c$ is a not a free parameter but is instead fixed to provide the desired number density of objects, in our case $\bar{n}_{\mathrm{gal }}^{\mathrm{ELG}} = 2.3 \times 10^{-4} h^{3} \mathrm{Mpc}^{-3}$ by matching $n_{\mathrm{gal}}^{\mathrm{HOD}}$ to Eq. \eqref{eq:ngal}. 
Similarly, for LRG, the HOD is parameterised by ($M_{1}$,$\sigma_{\mathrm{M}},\alpha, \kappa,f_{\mathrm{ic}}$), and $M_{\mathrm{min}}$ is fixed in order to obtain twice the mean observed density of the LRG sample (to reduce sample noise), where $\bar{n}_{\mathrm{gal }}^{\mathrm{LRG}} = 7.8 \times 10^{-5} h^{3} \mathrm{Mpc}^{-3}$ as in \cite{Zhai2017}. This is done by matching $n_{\mathrm{gal}}^{\mathrm{HOD}}$ to $\bar{n}_{\mathrm{gal }}^{\mathrm{target}} = 15.6 \times 10^{-5} h^{3} \mathrm{Mpc}^{-3}$ in Eq. \eqref{eq:ngal}.
We therefore have three free parameters in our ELG HOD model, $\theta \subset (A_s,M_{\mathrm{min}},\alpha)$, and five parameters, $\theta \subset (M_{1},\sigma_{\mathrm{M}},\alpha, \kappa,f_{\mathrm{ic}}$), with the LRG HOD model. 
In recent works \citep{Avila2020,Rocher2023a,Rocher2023b,Yuan2022}, it is common to enhance this description with extra parameters that describe deviation from the satellite NFW profile and from the velocity assignment of Eq. \eqref{eq:velocitysat} in order to properly model the anisotropic correlation function. Since we are only interested in describing projected quantities (see Sect. \ref{sec:fitprocedure}), we did not introduce these additional parameters. Indeed, the GGL signal, being a real space quantity, is insensitive to galaxy velocities, while redshift space distortion has a subdominant impact on projected clustering \citep{vanderbosh2013,Rocher2023a}. 
Eventually, the HOD model described in this section was calibrated against data to provide realistic mocks with clustering properties similar to those of the observations, as explained in Sect. \ref{sec:fitprocedure}.

\subsection{Extended halo occupation distribution including assembly bias} \label{sec:ABpara}
Since we aim to provide realistic galaxy mocks that are in agreement with observations, we extended our HOD framework with additional parameters to take into account secondary halo properties. This was motivated by studies such as \cite{Leauthaud2017} that found strong differences between the observed galaxy-galaxy lensing signal of CMASS galaxies and the one measured in the mock catalogues using the standard HOD model described above. 
Here, we followed the formalism described in \cite{Xu2021,Yuan2022}. The standard HOD framework was extended with four additional parameters, two depending on halo properties (specifically the concentration and shape of the halo) and two external properties, namely the local density and local shear (local density anisotropies). Next, we define in practice how we estimated these additional parameters.

First, we defined the halo concentration $c$ and shape $e$ as 
\begin{equation}
c=\frac{r_{\mathrm{vir }}}{r_{s}}, \quad e=\frac{1 - b/a}{1 + b/a},
\end{equation}
with $r_{\mathrm{vir }}$ as the virial radius, $r_{s}$ as the scale radius of the density profile, and $b/a$ as the ratio between the major and semi-major axes of the dark matter halo ellipsoid. These halo properties were directly extracted from the halo catalogues. The concentration was measured by fitting an NFW profile to each halo, while shapes were determined with the method of \cite{Allgood2006}.

To specify the halo environment within the simulation, we employed two different techniques. The first one consisted of finding for each halo all neighbouring halos (including sub-halos) beyond its virial radius but within $r_{\mathrm{max}}= 5 \, h^{-1}\mathrm{Mpc}$ of the halo centre. This was done with the Python package\texttt{scipy-Kdtree},\footnote{\url{https://scipy.org/}} which keeps track of pairs of halo per object. The $r_{\mathrm{max}}$ scale was chosen in \cite{Yuan2021}, as values of $r_{\mathrm{max}}$ in the range $ \approx 4-6 \, h^{-1}\mathrm{Mpc}$ were found to better reproduce the clustering of CMASS galaxies. In particular, it was shown by the authors that the goodness of fit depends on the scale at which the environment is defined (see Fig. 8 of \cite{Yuan2021}). After summing up the mass of all these neighbouring halos, the environmental parameter $f_{\mathrm{env}}$ could then be computed as
\begin{equation}
\label{eq:fenv_nrv}
f_{\mathrm{env}} = M_{\mathrm{env}} / \bar{M}_{\mathrm{env}}(M) - 1,
\end{equation}
where $\bar{M}_{\mathrm{env}}(M)$ corresponds to the mean environment mass within a given halo mass bin $dM$. This environmental parameter hence corresponds to the halo over-density in an annulus around each halo. We computed $\bar{M}_{\mathrm{env}}(M)$ in the logarithm mass range $[11,15.0] \, h^{-1}$Mpc in bins of dex $\der M = 0.1 \, h^{-1}$Mpc.

The second method consisted of painting the dark matter particle field onto a mesh in order to evaluate the over-density field $\delta(\textbf{x})$ \citep{Paranjape2018,Delgado2022,Hadzhiyska2023a}. The field $\delta$ was then smoothed with a Gaussian kernel of radius $R$ and interpolated at the position of each halo. This was done with the Python package \texttt{nbodykit}.\footnote{\url{https://nbodykit.readthedocs.io/en/latest/}} From the smoothed density field, we further defined the environmental shear, which quantifies the amount of local density anisotropy. First, we computed the tidal tensor field as 
\begin{equation}
T_{i j}(\mathbf{x})=\partial_i \partial_j \psi_R(\mathbf{x}),
\end{equation}
where the normalised gravitational potential $\psi_R(\mathbf{x})$ obeys the Poisson equation
\begin{equation}
\nabla^2 \psi_R(\mathbf{x})=\delta_R(\mathbf{x}).
\end{equation}
This was easily done in Fourier space, the expression for the tidal tensor being equal to 
\begin{equation}
T_{i j}(\mathbf{x})=\mathrm{FT}\left\{\left(k_i k_j / k^2\right) \delta_R(\mathbf{k})\right\},
\end{equation}
from which we determined the tidal shear $q_R^2$:
\begin{equation}
q_R^2 \equiv \frac{1}{2}\left|\left(\lambda_2-\lambda_1\right)^2+\left(\lambda_3-\lambda_1\right)^2+\left(\lambda_3-\lambda_2\right)^2\right|,
\end{equation}
where the $\lambda$ correspond to the eigenvalues of $T_{i j}$.
Different definitions of the smoothing radius have been used in the literature. \cite{Delgado2022} used a top hat filter of fixed size: $5 \, h^{-1}\mathrm{Mpc}$.\footnote{A top hat filter of size $R$ is equivalent to a Gaussian filter of size $R/\sqrt{5}$ (see \cite{Paranjape2018}).} \cite{Hadzhiyska2023a} estimated the density field with a Gaussian filter at different smoothing scales and selected for each halo the closest smoothing scale to the virial radius of the halo, rounding up. In \cite{Paranjape2018}, they found that the correlation between large-scale linear bias and local shear was maximum at scales of four to six times the virial radii. We therefore adopted two definitions for our smoothing scale. The first one is a constant smoothing of size R = $5/\sqrt{5} \approx 2.25 \, h^{-1}\mathrm{Mpc}$, from which we determined $q_R^2$ for each halo. Our second definition uses a smoothing scale of $2.25 r_{\text{vir}}$ for each halo, from which we determined the local density and local shear. The density field was smoothed from 1 to 5 $h^{-1}\mathrm{Mpc}$ with a step $\mathrm{d}x = 0.25 \, h^{-1}\mathrm{Mpc}$ and interpolated at the position of each halo to determine $\delta_R(\mathbf{x})$ for $R = 2.25 r_{\text{vir}}$. 

These secondary halo properties were used to modify the two mass parameters of the HOD model, $M_{\mathrm{min}}$ and $M_1$, as 
\begin{align} \label{eq:AB_definition}
\log M_{\mathrm{min}}^{\mathrm{mod}}= & \log M_{\mathrm{min}} + A_{\mathrm{cent}} f_A + B_{\mathrm{cent}} f_B \\
\label{eq:AB_definition2}
\log M_1^{\mathrm{mod}}= & \log M_1 +A_{\mathrm{sat}} f_A + B_{\mathrm{sat}} f_B, 
\end{align}
 where $f_A$ and $f_B$ correspond to the distributions of internal (either concentration or shape) and environmental (either shear or density) properties of the halos. This type of effective description that modifies central and satellite occupation based on secondary properties was also proposed in \cite{Hadzhiyska2023b}. The authors found that the simulation-predicted quantities $N_{\mathrm{cent}}(M,f_A,f_B)$ and $N_{\mathrm{sat}}(M,f_A,f_B)$ for each mass bin are well approximated by hyper-plane surfaces, which are nearly constant over mass. These distributions were normalised in the range $[-1,1]$ within the mass bins defined to compute the environmental factor $f_{\mathrm{env}}$ in Eq. \eqref{eq:fenv_nrv}. This was done by first taking the natural logarithm of each quantity (1+$\delta$ for the density) before scaling and normalising it around zero. Taking the logarithm of each quantity first has the advantage of keeping the shape of their distributions intact, as we are much less sensitive to outliers that squeeze the distribution after normalisation. We present these distributions in Appendix \ref{sec:appendixB}. The high volume of {\sc uchuu} enabled us to probe the assembly bias and environmental dependencies on halo occupation up to very high halo masses, $M =14.8 \, h^{-1}$Mpc, with more than 100 halos in the mass bin $[14.7,14.8]$ $h^{-1}$Mpc. Extra parameters in Eq. \eqref{eq:AB_definition} were fixed to zero above this threshold, as higher masses bins are not populated enough for a robust statistical description. 
 
 Our extended HOD framework therefore has seven parameters for an ELG HOD $\theta \subset (A_s,M_{\mathrm{min}},\alpha,A_{\mathrm{cent}},B_{\mathrm{cent}},A_{\mathrm{sat}},B_{\mathrm{sat}})$ and nine parameters $\theta \subset (M_{1},\sigma_{\mathrm{M}},\alpha, \kappa,f_{\mathrm{ic}},A_{\mathrm{cent}},B_{\mathrm{cent}},A_{\mathrm{sat}},B_{\mathrm{sat}}$) for an LRG HOD. In the following, we test different combinations in Eqs. \eqref{eq:AB_definition}-\eqref{eq:AB_definition2}, between internal halo properties $f_A$ (concentration and shape) and the local environment $f_B$ (density and shear).
 
\subsection{Fitting procedure} \label{sec:fitprocedure}
Our fitting method uses the HOD model described above to populate dark matter halos with galaxies for which clustering and lensing statistics are measured and compared to the ones of the data. Here, we are interested in the projected correlation function, which is almost insensitive to the effect of redshift space distortions at small scales. We computed the galaxy anisotropic two-point function $\xi\left(r_p, \pi\right)$ as a function of galaxy pair separation along $(\pi)$ and transverse to the line of sight (l.o.s.;$r_p$). Integrating over the l.o.s. yielded 
\begin{equation}
w_p\left(r_p\right)=2 \int_0^{\pi_{\max }} \xi\left(r_p, \pi\right) \der \pi\;.
\end{equation}
We used the Dark Energy Spectroscopic Instrument \citep[DESI,][]{DESI2016} wrapper, \texttt{pycorr},\footnote{\url{https://py2pcf.readthedocs.io/en/latest/index.html}} around the \texttt{corrfunc} package \citep{Corrfunc} to compute $\xi\left(r_p, \pi\right)$ with the natural estimator, which normalises galaxy pair counts to analytical estimation of random pairs in a periodic box. For the data, we used the minimum variance  \citet{landy93} estimator
\begin{equation} \label{eq:LS}
\xi(r_p,\pi) = \frac{DD(r_p,\pi) - 2DR(r_p,\pi) + RR(r_p,\pi)}{RR(r_p,\pi)},
\end{equation}
with $R$ as a random catalogue with the same angular and redshift distribution as the data. We applied to each galaxy and random catalogue the total weight $w=w_{\mathrm{noz}} w_{\mathrm{cp}} w_{\mathrm{syst}} w_{\mathrm{FKP}}$ (see \cite{ross20} for an in-depth definition of each individual weight). Briefly, $w_{\mathrm{noz}}$ corrects for redshift failures, $w_{\mathrm{syst}}$ corrects for spurious fluctuation of target density due to observational systematic effects, FKP weights \citep{Feldman1994} are optimal weights used to reduce variance in clustering measurements, and close-pair weights $w_{\mathrm{cp}}$ correct for fibre collisions. Correcting for fibre collision is an important matter in our case, as we want to extract information from the one-halo regime in order to constrain the satellite distribution in our HOD model. Traditionally, these weights are applied by up-weighting observed galaxies in a collision group. \\
\begin{figure*}
    \centering    
    \includegraphics[width=15cm]{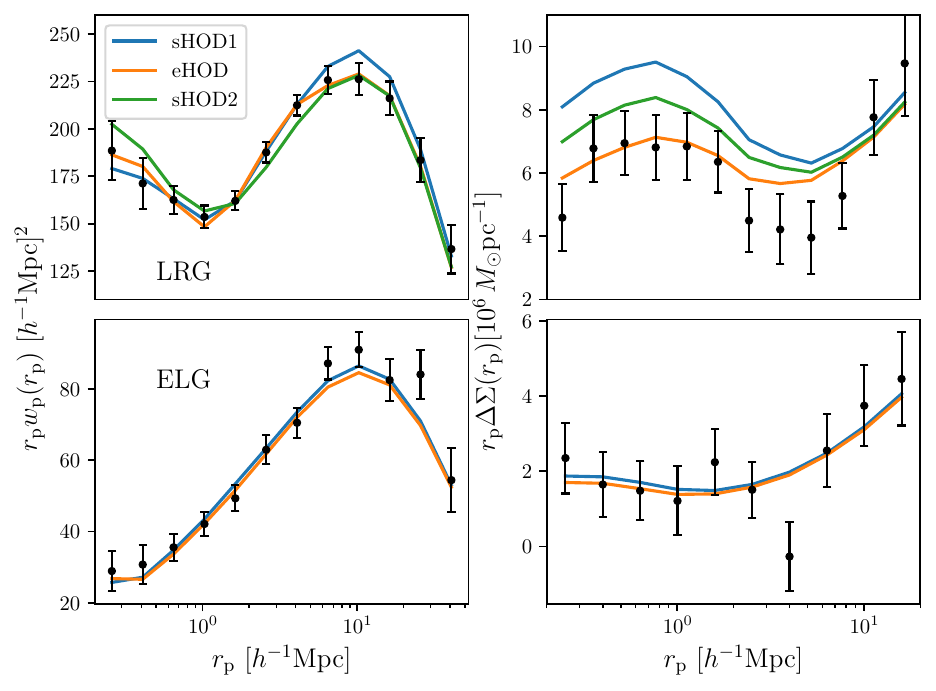}
    \caption{Best-fit HOD models for LRG (top) and ELG (bottom). The subscript `s/e' denotes the standard and extended HOD model. The extended model here corresponds to our fiducial configuration, where the concentration and an annular definition of $\delta$ are used as additional parameters. The points with error bars correspond to the measured $w_p$ (left) and the $\Delta \Sigma$ (right) signals. The blue lines (sHOD1) correspond to a fit to the projected correlation function $w_p$ only, while the green (SHOD2) and orange lines (eHOD) correspond to standard and extended HOD fits including the measurement of $\Delta \Sigma$ in the fitting pipeline. The extended HOD model provides a good description of both the clustering and lensing signal.
    We did not perform any standard HOD fit to the combined $w_p$ + $\Delta \Sigma$ vector for ELGs (sHOD2), as the standard HOD fit to $w_p$ only already agrees well with the $\Delta \Sigma$ measurement.}
    \label{fig:main_LRG_plot}  
\end{figure*}
\begin{figure}
    \centering
    \includegraphics[width=\columnwidth]{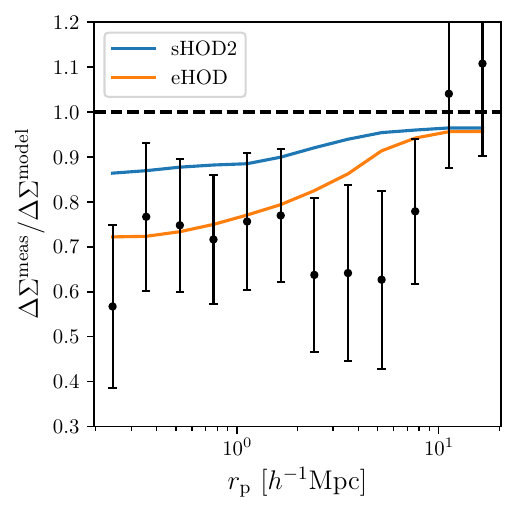}
    \caption{Difference between the observed galaxy-galaxy lensing signal  (points with error bars) of eBOSS LRGs with the prediction of a standard HOD (sHOD) determined by fitting the projected correlation function $w_{\mathrm{p}}$ only. The continuous line in blue represents an sHOD fit to $w_{\mathrm{p}} + \Delta \Sigma$, while the orange line corresponds to the extended HOD model. 
      }
    \label{fig:diff_LRG}  
\end{figure}
While this method provides a good correction scheme at large scales, which is enough for the study of galaxy anisotropic clustering, it fails on smaller scales, and we therefore needed another method to correct for fibre collisions. In particular, we used the pairwise-inverse probability (PIP) and angular upweighting (ANG) weights described in \cite{Mohammad2020}. These weights were shown to properly account for fibre collisions at small separations. However, PIP weights were only produced for eBOSS galaxies, and we therefore only applied ANG weights to the combined sample (including the $cp$ weights), as they provide sufficient correction below 1 $h^{-1}$Mpc. 
For $w_p(r_p)$, we used 13 logarithmic bins in $r_p$ between 0.2 and 50 $h^{-1}$Mpc, with $\pi_{\mathrm{max}}$ = 80 $h^{-1}$Mpc.
In addition to the clustering, we also fit the GGL signal, similar to the work of \cite{Contreras2023}, observed with eBOSS lenses around DES sources in order to further constrain the assembly bias parameters. The measured GGL differential excess surface density is defined as
\begin{equation}
\Delta \Sigma_{\mathrm{gm}}(r)=\bar{\Sigma}_{\mathrm{gm}}(r)-\Sigma_{\mathrm{gm}}(r),
\end{equation}
where the mean projected surface density is defined as
\begin{equation}
\bar{\Sigma}_{\mathrm{gm}}(r)=\frac{2}{r^2} \int_0^r \Sigma_{\mathrm{gm}}(r') r' \mathrm{d} r'.
\end{equation}
The projected surface density $\Sigma_{\mathrm{gm}}(r)$ can be written as a function of the galaxy-matter cross-correlation as \citep{Guzik2001,delatorre2017,Jullo2019} 
\begin{equation}
\label{eq:projected_surface}
\Sigma_{\mathrm{gm}}(r)=\rho_{\mathrm{m}} \int_{-\infty}^{\infty} \xi_{\mathrm{gm}}\left(\sqrt{r^2+\chi^2}\right) \mathrm{d} \chi,
\end{equation}
with the mean matter density $\rho_{\mathrm{m}}=\Omega_{\mathrm{m}}(z) \rho_{\mathrm{c}}(z)=3 \Omega_{\mathrm{m}} H_0^2 / 8 \pi G$ constant in redshift in co-moving coordinates. The galaxy-matter correlation $\xi_{\mathrm{gm}}$ function was measured in the range [$5 \times 10^{-3}, 150$] $h^{-1}$Mpc with 41 logarithm bins by cross-correlating galaxy mock catalogues with the dark matter particles of the simulation, sub-sampled by a factor of 2000. With this sub-sampling factor, we could accurately predict $\Delta \Sigma$ (we omit the $\mathrm{gm}$ subscript in the following parts of the paper) as in \cite{Contreras2023b}, whose authors claimed that sub-sampling the matter field by a factor $\sim$ 0.0003 ensures    
robust measurement of $\Delta \Sigma$ with sub-percent precision across all the ranges considered in their analysis. We checked this statement with the simulation used in this work. The measurements of $\Delta \Sigma$ performed with particles sub-sampled by a factor 0.005\footnote{This is the maximum number of particles that can be extracted from the simulation.} or 0.0005 are the same.
We then estimated $\Sigma(r)$ by integrating Eq. \eqref{eq:projected_surface} up to 100 $h^{-1}$Mpc. The estimation of $\Delta \Sigma$ was then compared to the one observed in the data, which is computed as \citep{Jullo2019}
\begin{equation}
\begin{aligned} \label{eq:LS_DS}
\Delta \Sigma \left(r_{\mathrm{p}}\right)= & \frac{\sum_{\mathrm{l}, \mathrm{s}}^{N_{\mathrm{ls}}} \sum_{\mathrm{cr}}\left(z_1, z_{\mathrm{s}}\right) w_{\mathrm{l}, \mathrm{s}} \epsilon_{+}\left(r_{\mathrm{p}}\right)}{\sum_{\mathrm{l}, \mathrm{s}}^{N_{\mathrm{l}, \mathrm{s}}} w_{\mathrm{l}, \mathrm{s}}} \\
& -\frac{\sum_{\mathrm{r}, \mathrm{s}}^{N_{\mathrm{r}, \mathrm{s}}} \sum_{\mathrm{cr}}\left(z_{\mathrm{r}}, z_{\mathrm{s}}\right) w_{\mathrm{r}, \mathrm{s}} \epsilon_{+}\left(r_{\mathrm{p}}\right)}{\sum_{\mathrm{r}, \mathrm{s}}^{N_{\mathrm{r}, \mathrm{s}}} w_{\mathrm{r}, \mathrm{s}}},
\end{aligned}
\end{equation}
\begin{table*}
  \caption{Best-fit parameters of the LRG HOD for all configurations defined in the main text.}
  \begin{tabular}{lp{1.5cm}p{1.5cm}p{1.5cm}p{1.5cm}p{1.5cm}p{1.5cm}p{1.5cm}p{1.5cm}}
\hline
\hline
Method  &  sHOD1 & sHOD2 & eHOD $\delta_{\mathrm{vir}}$,c & eHOD $T^{ij}_{\mathrm{vir}}$,c & eHOD $T^{ij}_{\mathrm{vir}}$, b/a &  eHOD $\delta_{\mathrm{annulus}}$,c & eHOD $\delta_{\mathrm{annulus}}$,b/a & eHOD $T^{ij}_{5\mathrm{Mpc}}$ ,c \\ 
\hline 
\noalign{\vskip 1pt}
   $ M_{\mathrm{min}}$ & 13.24 & 12.81 & 12.95 & 13.23 & 12.93 & 13.03 & 13.11 & 13.26  \\
   $M_{1}$ & 14.18 & 13.84 & 13.91 & 13.80 & 13.72 & 13.76 & 13.76 & 13.87 \\
   $\sigma$ & 0.54 & 0.11 & 0.44 & 0.74 & 0.53 & 0.58 & 0.67 & 0.72 \\
   $\alpha$ & 1.20 & 1.36 & 0.91 & 0.93 & 1.05 & 0.93 & 0.90 & 0.97 \\
   $\kappa$ & 0.39 & 0.36 & 0.64 & 0.48 & 0.39 & 0.56 & 0.51 & 0.46 \\
   $f_{\mathrm{ic}}$ &  0.64 & 0.28 & 0.32 & 0.38 & 0.25 & 0.30 & 0.32 & 0.43 \\
   $A_{\mathrm{cent}}$ &&& 0.11 & - 0.01 & 0.10 & 0.18 &  0.20 & 0.15 \\
   $B_{\mathrm{cent}}$ &&& - 0.27 & -0.27 & -0.24 & -0.25 & -0.25 & -0.18 \\
   $A_{\mathrm{sat}}$ &&& 0.01 & -0.15 & -0.05 & -0.06 & -0.08 & -0.02 \\
   $B_{\mathrm{sat}}$ &&& 0.00 & -0.04 & -0.10 & 0.01 & 0.02 & 0.01 \\ 
   $f_{\mathrm{sat}} \%$& 8.0 & 14.1 & 13.3 & 12.8 & 16.6 & 14.8 & 14.6 & 12.0 \\
   $\chi^{2}_{r}$ & 1.05(2.42) & 1.93 & 1.43 & 1.11 & 1.06 & 1.08 & 1.09 & 1.62 \\
   \hline
\end{tabular}
\tablefoot{The baseline eHOD configuration in this work corresponds to the case where the concentration and an annular definition of $\delta$ are used as additional parameters. The standard model sHOD1 only fit $w_p$, while every other model fit the combined $w_p$ + $\Delta \Sigma$ data vector. For the standard HOD model (sHOD1), we show in parenthesis the reduced chi-square including $\Delta \Sigma$. The full posterior distributions corresponding to these bestfit parameters are presented in Appendix \ref{sec:appendixC}.}
\label{tab:bestfit_values_HOD_LRG}
\end{table*}
where $z_{\mathrm{s}}$ and $z_{\mathrm{l}}$ correspond to the source and lens redshift, respectively, and $\epsilon_+$ represents the tangential
component of a source ellipticity around a lense. The subtraction around random lenses reduces the contribution of lensing systematics and leads to a reduced variance on small scales \citep{Singh2017}. The critical density $\Sigma_{\mathrm{cr}}$ is defined as 
\begin{equation}
\Sigma_{\mathrm{cr}}\left(z_1, z_{\mathrm{s}}\right)=\frac{c^2}{4 \pi G\left(1+z_{\mathrm{s}}\right)^2} \frac{D_{\mathrm{S}}}{D_{\mathrm{LS}} D_{\mathrm{L}}},
\end{equation}
where $D_{\mathrm{S}},D_{\mathrm{LS}}, D_{\mathrm{L}}$ are the observer-source, lens-source, and observer-lens angular diameter distances.
We measured $\Delta \Sigma (r_p)$ with 13 (11 for the ELG) logarithmic bins in the range between 0.2 and 20  $h^{-1}$Mpc with the python package \texttt{dsigma}\footnote{\url{https://dsigma.readthedocs.io/en/stable/index.html}} \citep{Lange2022}. We used \texttt{Metacalibration} shape measurements and corrected for potential systematic effects induced by this method. We used for photometric redshifts of source galaxies the ones estimated with the Directional Neighbourhood Fitting (\texttt{DNF}) algorithm \citep{Devicente2016}. Errors in photo-$z$ measurements can lead to bias in the estimation of $\Delta \Sigma$. We matched DES galaxies with galaxies observed spectroscopically with BOSS and eBOSS. While these samples are not representative of DES galaxies, they can however be used to 
obtain a rough estimate of the bias introduced by the photo-$z$. We found that photo-$z$ leads to an underestimation of the observed signal by approximately $3-5\%$ at all scales. Since this effect is sub-dominant compared to the S/N of our measurements, we did not include it. We also did not include any Boost factor correction \citep{Miyatake2015}, as we have found that it has nearly no impact on our measurements, even at the smallest scale. 

When computing the correlation functions for the data in Eqs \eqref{eq:LS} and \eqref{eq:LS_DS}, we assumed the {\sc uchuu} cosmology to compute radial distances from redshift. We then compared the HOD clustering and lensing signals to the observed ones in eBOSS by means of a likelihood analysis:
\begin{equation} \label{eq:likelihood}
-2 \ln \mathcal{L}(\theta)=\sum_{i, j}^{N_p} \vec{\Delta}_i(\theta) \hat{\Psi}_{i j} \vec{\Delta}_j(\theta),
\end{equation}
where $\theta$ is the set of HOD parameters, $\Delta$ is the data-model difference vector, $N_p$ is the total number of data points, and $\hat{\Psi}_{i j}$ is the precision matrix, the inverse of the covariance matrix. The covariance matrix is determined given the observational measurements by jackknife (JK) sub-sampling with $N = 100$ patches. Since all ELGs are within the DES footprint, we could use the same JK regions for the clustering and lensing measurements in order to estimate the cross-correlation between $w_{\mathrm{p}}$ and $\Delta \Sigma$. For the LRG, only about a third of the sample is within DES footprint. We decided to use the whole sample to estimate $w_{\mathrm{p}}$ in order to get a sufficient S/N measurement. Since $\Delta \Sigma$ is only measured within the overlapping footprint, the cross-covariance cannot be estimated through the jackknife method, and we assumed it to be zero. We checked with the ELG sample that this assumption does not change the results significantly: the jackknife cross-covariance is mainly dominated by noise. 

Our fitting procedure corresponds to the one described in \citep{Rocher2023a,Rocher2023b}.  The parameter space is first sampled with a quasi-uniform grid, which serves as a training sample for a Gaussian process regressor and technical details. Here, we used the Gaussian process implemented in \texttt{sklearn}.\footnote{\url{https://scikit-learn.org/stable/}} For each sampled point in the parameter space, the clustering and lensing signals are averaged over ten realisations to reduce stochastic noise from halo occupation. We then repeated, iteratively, the following procedure: We used a Markov chain Monte Carlo (MCMC) sampler, here \texttt{emcee}\footnote{\url{https://emcee.readthedocs.io/}} \citep{Foreman2013}, to sample our parameter space given our surrogate model generated by the Gaussian process regressor. Specifically, we ran 50 chains of 5,000 points in parallel, with the first 500
points being discarded in each chain.
 We then randomly picked three points within our final MCMC chains from which we computed the likelihood. We then added these points to the initial training sample and re-trained our Gaussian process. In \cite{Rocher2023a}, they found that iterating up to $N=800$ from a training sample of $N_{\text{ini}}\approx 1000$ was enough to get robust contours. These findings are highly dependent on the parameter space one wants to explore and the dynamic range of the likelihood. While we do find the same conclusion for an ELG HOD, the degree of degeneracy of an LRG HOD calls for a higher number of iterations. We therefore used $N_{\text{ini}}= 1000$ (resp. 2000) and $N=1000$ (resp. 2000) for the ELG (resp. LRG) fit. We checked that our results are robust to these choices. Additionally, we refer the reader to \citep{Rocher2023a} for more details on the performance of the Gaussian process emulator. Briefly, the authors compared their results with the \textsc{abacus} HOD pipeline \citep{Yuan2021} and found strong agreements between the two methods. The main difference between the implementation of \cite{Rocher2023a} and ours is that the authors only picked one point in each chain for each iteration, while we decided to pick three per iteration. The reason for this decision is computation speed. We found that for a high number of points, the Gaussian process regression takes a longer time to compute than three averaged HOD realisations. The key element is to have a high enough number of points in the region where the likelihood is minimal to have robust contour estimation with the Gaussian process.
\section{Halo occupation distribution fit} \label{sec:hodfit}
\begin{figure*}
    \sidecaption
    \includegraphics[width=12cm]{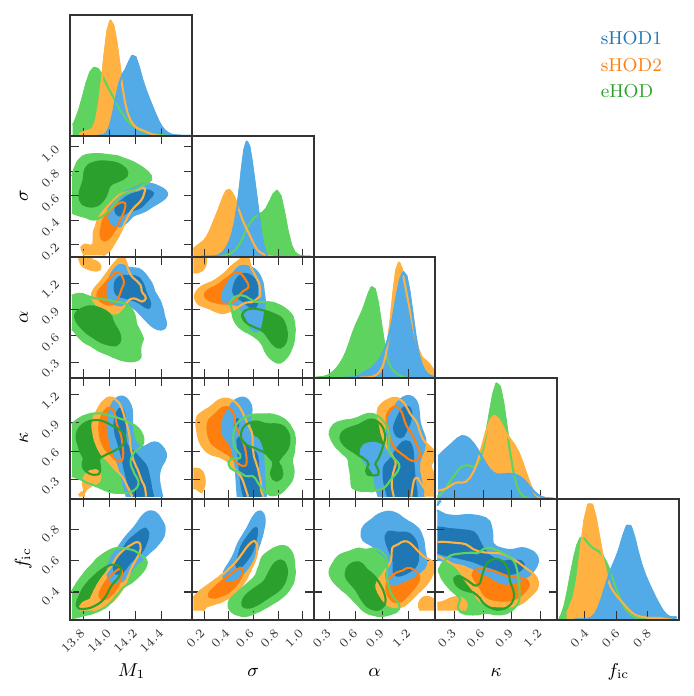}
    \caption{Posterior contours of the sHOD1 (in blue), sHOD2 (in orange), and eHOD (in green) in our baseline configuration (including concentration and annular density as additional parameters) for the LRG sample.
    One can observe that the extra parameters contained in the eHOD model have a strong impact on the posterior distributions.}
    \label{fig:contours_LRG}  
\end{figure*}

\begin{figure}
    \centering    
    \includegraphics[width=\columnwidth]{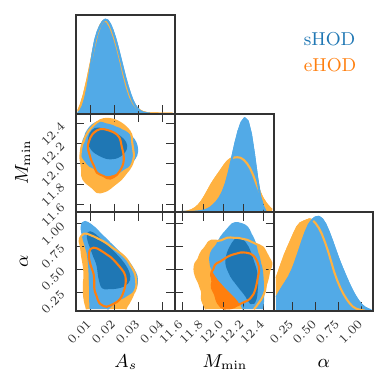}
    \caption{Posterior contours for the sHOD (in blue) and eHOD (in orange) models for the ELG sample. Taking into account assembly bias parameters has little impact on halo occupation. 
      }
    \label{fig:full_posterior_ELG_HOD}  
\end{figure}

\begin{figure}
    \centering    
    \includegraphics[width=\columnwidth]{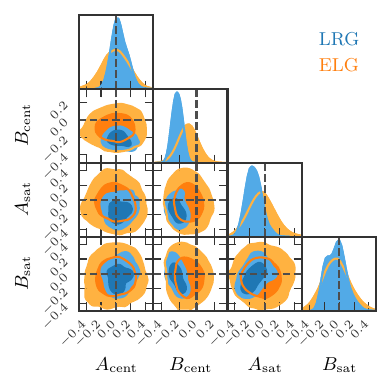}
    \caption{Posterior contours of the assembly bias and environmental parameters in our baseline configuration (see Sect. \ref{sec:result_LRG}). For the LRG, the negative value of $B_{\mathrm{cent}}$, detected at $3\sigma$, indicates that LRG centrals most likely occupy a dense environment. A negative value of  $A_{\mathrm{sat}}$ at the 1$\sigma$ level indicates that satellite galaxies could be preferentially located in more concentrated halos than average.  For the ELGs, all parameters are consistent with zero at the $1\sigma$ level, suggesting that ELGs do not exhibit significant assembly bias or environmental dependencies.
      }
    \label{fig:contours_AB}  
\end{figure}

\begin{figure*}
    \centering    
    \includegraphics[width=15cm]{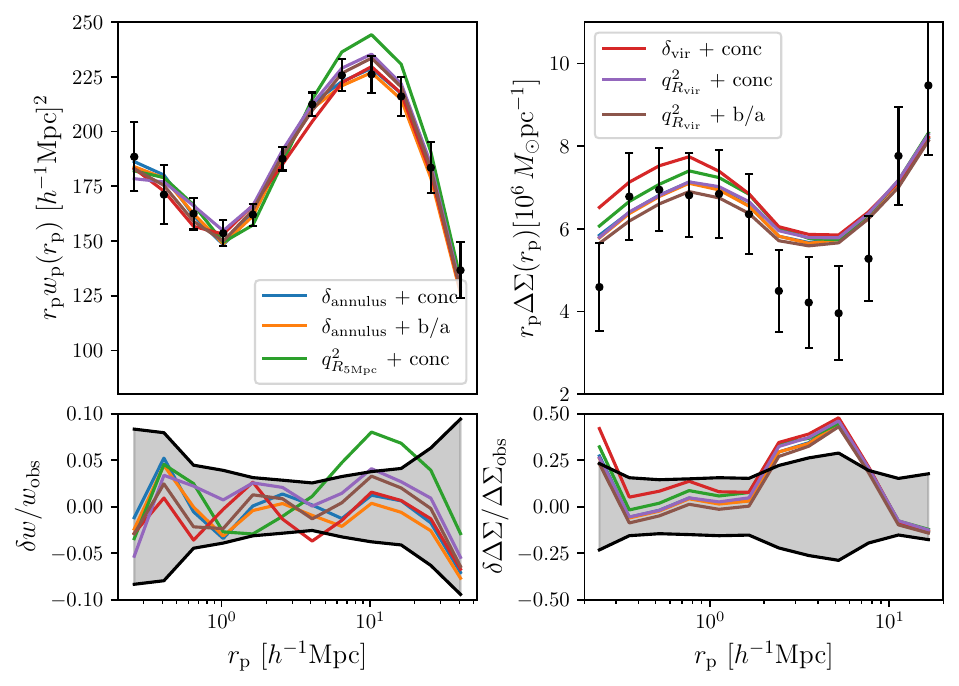}
    \caption{Result for different combinations of assembly bias (concentration and shape) and local environments (density and shear) parameters. Top panels present the measurements, while the bottom panels show the relative differences. Nearly all the models provide similar best-fitting clustering predictions, lensing predictions, and goodness of fit results except the shear measured with a Gaussian Kernel of size 2.25 $h^{-1}$Mpc, which has a slightly worse goodness of fit.
      }
    \label{fig:main_LRG_plot_v2}  
\end{figure*}

\subsection{Luminous red galaxies} 
\label{sec:result_LRG}
As a baseline configuration for the extended HOD (eHOD) model, we used the concentration as a proxy of assembly bias and the local density, described as an annular definition of the density field $\delta$, following the formalism of \cite{Yuan2022}. We emphasize that the annular definition of local density might be prone to sub-halo identification and thus simulation resolution. However, given the properties of the {\sc uchuu} simulation (see Sect. \ref{sec:uchuu}), we did not consider this potential source of bias in our analysis. We present in Fig. \ref{fig:main_LRG_plot} the best-fit results obtained with the standard (s) and extended (e) HOD models. The standard model sHOD2 and the extended model both fit the combined $w_{p}$, $\Delta \Sigma$ data vector, while the sHOD1 model only fits $w_{p}$. Best-fit values are given in Table \ref{tab:bestfit_values_HOD_LRG}, while full posterior distributions and prior information can be found in Appendix \ref{sec:appendixC}. For the sHOD1, one can see that the model over-predicts the signal at scales around 10 $h^{-1} \mathrm{Mpc}$ while obtaining a satellite fraction of $8 \%$, slightly lower than what was obtained in \cite{Alam2020,Yuan2022} and $50 \%$ below the best-fit value of \cite{Zhai2017}. We found that this difference was due to noise in the jackknife covariance estimate of $w_{\mathrm{p}}$. When we performed a fit only considering the diagonal elements, we obtained a satellite fraction of $11 \%$, the model matches well the peak at  10 $h^{-1} \mathrm{Mpc}$ but in turn provides a poor fit at scales $1-3 h^{-1} \mathrm{Mpc}$, similar to the observed fit for the sHOD2 case and an issue in the modelling also reported in \cite{Zhai2017}. The introduction of secondary HOD parameters enabled us to provide a better description of the small and intermediate scales of $w_{\mathrm{p}}$ while also yielding a more accurate description of $\Delta \Sigma$, decreasing the reduced chi-square from $\chi^2_r = 1.93$ down to $\chi^2_r = 1.08$, an improvement mainly due to a better description of $\Delta \Sigma$ on the scales below $3 \,h^{-1}$Mpc. 

Figure \ref{fig:diff_LRG} displays the ratio of $\Delta \Sigma$ measurements in the data and as measured in the mocks generated by sHOD2 and eHOD models (calibrated on both $w_p$ and $\Delta \Sigma$ measurements), compared to the $\Delta \Sigma$ measurement as measured in the mock generated by the sHOD1 (calibrated only against the projected clustering $w_p$). The standard model (dashed line) over-predicts the GGL signal by $30-40 \%$ at scales below 10$h^{-1}$ Mpc, while a combined fit (blue solid line) poorly alleviates this issue, with a slight improvement of only $ \approx 5-15 \%$. With the eHOD model, we observed an improvement of 20$\%$ on scales between $2-10 h^{-1}$ Mpc and a really good match to the data at lower scales. This better goodness of fit is mainly achieved by incorporating a dependence on the local density environment \citep{Yuan2021,Hadzhiyska2023b}. Thus, adding secondary halo parameters to the HOD models helps reproduce the lensing signal at small scales and gives a partial answer to the lensing-is-low problem.
Indeed, baryonic feedback, not considered in this work, smooths out the matter distribution, also contributing to a decrease of $\Delta \Sigma$ \citep{Amodeo2021}.

We present in Fig. \ref{fig:contours_LRG} the posterior distributions of each HOD fit. The better description of $\Delta \Sigma$ with the sHOD2 was achieved by lowering the mass threshold $M_1$ and $M_{\textrm{min}}$ (see also Table \ref{tab:bestfit_values_HOD_LRG}), above which central and satellite galaxies occupy dark matter halos. We observed the same trend for the eHOD, interestingly with a higher best-fit value for $M_{\textrm{min}}$ compared to the sHOD2 case. To maintain a similar galaxy bias between the eHOD and sHOD1 models, galaxies in lower mass halos should therefore be located in higher density peaks. This is what we obtained in the posterior distributions of Fig. \ref{fig:contours_AB}. A negative value of $B_{\mathrm{cent}}$ implies that halos located in denser environments are more likely to be populated, as a decrease of $M_{\mathrm{min}}^{\mathrm{mod}}$ in Eq. \eqref{eq:AB_definition} increases the probability of populating such halos, a result also found for CMASS galaxies \citep{Yuan2022} and in hydrodynamical simulations \citep{Artale2018}.

In recent studies, halo occupation was found to be strongly dependent on formation time \citep{Zehavi2018,Artale2018} and concentration \citep{Bose2019,Delgado2022}, as both variables are correlated \citep{Wechsler2002}, although in a non-trivial way \citep{Gao2007}. The conclusion from these analyses was that at low mass, central galaxies are more likely to be hosted by more concentrated, early-formed halos. This trend reverses at higher mass due to the contributions of satellite galaxies.
However, given our measurements, the only parameter that is well constrained is $B_{\mathrm{cent}}$, being non-zero at the 3$\sigma$ level, while all the other parameters are consistent with zero at the 1$\sigma$ level. Hence, based on our analysis, we cannot draw a proper conclusion on the impact of assembly bias on halo occupation. Similar to \cite{Yuan2022}, we also find that including secondary assembly bias parameters (or just fitting $\Delta \Sigma$) does increase the satellite fraction. This can be explained by the fact that central-particle correlation is dominant on small scales \citep{Leauthaud2017}. Thus, increasing the satellite fraction will lower the central contribution and the low scales $\Delta \Sigma$ signal. For completeness, we present the full posterior distribution of the extended HOD in Appendix \ref{sec:appendixC}.
In Sect. \ref{sec:standardHOD}, we made two approximations that might impact the result present in this section, mainly the choice of number density and an assumption of a simple NFW profile. We show in Appendix \ref{sec:appendixA} that our statistically significant result, the environmental dependence on halo occupation, is robust with respect to our fiducial choices. However, we did not consider other physical effects that impact our GGL signal. 

As demonstrated in \cite{2023MNRAS.521..937C}, satellite segregation and baryonic feedback are the two main sources, after assembly bias, that create the lensing-is-low problem. Clearly, including these effects would modify the interpretation of our results. Study of these effects is out of the scope of this paper, as our main goal is to investigate how small-scale variations of $\Delta \Sigma$ can impact cosmological analysis of anisotropic clustering plus lensing signal (see Sect. \ref{sec:cosmofit}).
\begin{table}
\caption{Posterior and best-fit results for the ELG HOD.} 
\begin{tabular}{lcc}
\hline
\hline
Method  &  sHOD1 & eHOD \\
\hline
\noalign{\vskip 1pt}
$ A_s$ [0.002,0.02] & $0.017^{+0.006}_{-0.005}$(0.015) & $0.017^{+0.006}_{-0.005}$(0.015)\\
$ M_{\mathrm{min}}$ [11.5,12.5] & $12.19^{+0.10}_{-0.12}$ (12.28)& $12.11^{+0.17}_{-0.19}$(12.21) \\
$\sigma$ [0.1,1.2] & $0.52^{+0.23}_{-0.23}$(0.41) &  $0.42^{+0.22}_{-0.23}$(0.28)\\
$A_{\mathrm{cent}}$ [-0.5,0.5] &  & $-0.01^{+0.19}_{-0.22}$ (-0.08)\\
$B_{\mathrm{cent}}$ [-0.5,0.5] &  & $-0.09^{+0.11}_{-0.19}$  (-0.04)\\
$A_{\mathrm{sat}}$ [-0.5,0.5] &   & $-0.04^{+0.19}_{-0.18}$ (0.02)\\
$B_{\mathrm{cent}}$ [-0.5,0.5] &   & $-0.04^{+0.20}_{-0.12}$(-0.08)\\
\hline
\noalign{\vskip 1pt}
$A_c$ [0.01] & $0.034^{+0.012}_{-0.011}$ (0.047) & $0.028^{+0.017}_{-0.012}$ (0.038)  \\
$f_{\mathrm{sat}} \%$& 10.4 & 13.3 \\
$\chi^2_r$ & 1.13(1.22) & 1.40
\end{tabular}
\tablefoot{We show here the 16, 50, and 84 percentiles of the posterior distribution. The first column presents the prior ranges for each parameter. The two other columns show the 16, 50, and 84 percentiles of the posterior distributions, with the best-fit values in parenthesis defined as the minimum of the $\chi^2$ of the MCMC chains for both sHOD and eHOD models.
The initial value of $A_c$ is fixed to 0.01, meaning that $A_s$ is allowed to vary in the range [0.2,20]$A_c$. Other columns present the rescaled values of $A_c$ and $A_s$ in order to match the ELGs number density. For the sHOD1, we show in parentheses the reduced chi-square including $\Delta \Sigma$.}
\label{tab:result_HOD_ELG}
\end{table}  

\subsection{Emission line galaxies} 
The best-fit results for ELGs are represented in the bottom panel of Fig. \ref{fig:main_LRG_plot}, while the posterior contours are shown in Fig. \ref{fig:full_posterior_ELG_HOD} with the corresponding constraints in Table \ref{tab:result_HOD_ELG}. We did not observe significant differences between the standard and the extended HOD framework when considering either the best-fit model or the full posterior distribution. Both models provide a similar goodness of fit and reproduce the observations, with a $\chi^2$ = 23.1 and 22.0 respectively for the sHOD and eHOD models, resulting in a worse reduced $\chi^2$ for the eHOD, as the four additional free parameters do not significantly improve the fit. From the posterior distributions in Fig. \ref{fig:contours_AB}, one can see that all the secondary parameters are consistent with zero at the $1 \sigma$ level. 
In the literature, it was shown that ELGs are preferentially located in denser regions than average \citep{Artale2018,Delgado2022} and in more concentrated halos \citep{Zehavi2018,Artale2018,Rocher2023b}. This is in agreement with our negative best-fit values for $A_{\mathrm{cent}}$ and $B_{\mathrm{cent}}$, even though we cannot derive more decisive conclusions from our analysis.

In addition, we found a satellite fraction of $10.4 \%$ for the sHOD, significantly lower than that reported in \cite{Avila2020}, $f_{\mathrm{sat}}= 22\%$. 
However, for ELGs, the mean satellite occupation is independent of that of the centrals, which means that an ELG satellite can orbit around a non-ELG central. Thus, it only contributes to the two-halo term of the ELG auto-correlation function, affecting the amplitude of the linear bias of the galaxy sample. The lower satellite fraction (compared to \cite{Avila2020}) can be explained by the fact that our mean value for $M_{\mathrm{min}} = 12.19$ is larger than that reported in \cite{Avila2020}, $M_{\mathrm{min}} = 11.71$ and that we allowed $\alpha$ to vary, resulting in a different satellite occupation over halo mass. In \cite{Rocher2023b}, when they include strict conformity, that is, satellite galaxies can only populate the halo in the presence of a central ELG, they report a very small fraction of satellite of $\approx 2-3 \%$. For DESI ELG, strict conformity is needed in order to provide a positive value for $\alpha$, which is predicted by semi-analytical models \citep{GP2018Fa,GonzalezPerez2020}.

\subsection{Dependence on parameter choices} \label{sec:diffHOD}
We present in Fig. \ref{fig:main_LRG_plot_v2} multiple eHOD best-fit models for different configurations. We tested as a proxy for the environment either the local shear or the local density, combined with internal properties, that is, concentration or shape of the dark matter halos. As a reminder, the baseline configuration corresponds to the case where we apply to Eq. \eqref{eq:AB_definition} $f_A$ = concentration and $f_B = \delta_{\mathrm{annulus}}$, with $ \delta_{\mathrm{annulus}}$ calculated by summing the mass of neighbouring halos. Estimating the local density field with this definition provides a better description of the observed LRG signal, with a $\chi_\mathrm{r}^2$ improved by $\Delta \chi^2_{\mathrm{r}}$ = -1.39, compared to the case where the density field is estimated on a mesh with a Gaussian smoothing kernel of size = $2.25 r_{\mathrm{vir}}$, with $\Delta \chi^2_{\mathrm{r}}$ = -0.98. These results are in agreement with those of \cite{Yuan2022}, who found that estimating the density field with a Gaussian kernel of a fixed size of 3 $h^{-1}$ Mpc yields no significant improvement compared to the sHOD. 
Instead of the concentration of the dark matter halo, using its shape as a proxy for assembly bias provides nearly indistinguishable best-fit model results. The resulting constraints, presented in Table \ref{tab:bestfit_values_HOD_LRG} and in Appendix \ref{sec:appendixC}, suggest that red galaxies preferentially occupy more ellipsoidal halos than average, as the best-fitting value $A_{\mathrm{cent}}$ is positive when either considering shear or density as external property. The two extra configurations including the local density are in agreement with the baseline configuration, which is in favour of a negative value for $B_{\mathrm{cent}}$ such that galaxies preferentially occupy denser environments. 

Using the local shear instead of the local density also provides good $\chi_{\mathrm{r}}^{2}$ values when the tidal tensor is computed with a Gaussian kernel of size 2.25 $r_{\mathrm{vir}}$. The best $\chi_\mathrm{r}^2$ value is obtained when combining the local shear with the halo shape, $\chi_\mathrm{r}^2$ = 1.06. It reproduces the small scales of $\Delta \Sigma$ better but gives a slightly worse fit around the peak of $w_\mathrm{p}$ observed at $10 h^{-1}$ Mpc.
Surprisingly, using the local shear with a fixed Gaussian smoothing of size 2.25 $h^{-1}$Mpc provides the worst $\chi_{\mathrm{r}}^2$ value out of all configurations. This demonstrates that local density anisotropy should be considered at halo scales to retrieve the maximum amount of information, as explained in \cite{Paranjape2018}. Our results suggest that galaxies preferentially occupy more anisotropic environments, in agreement with \cite{Delgado2022}, who found this correlation for low-mass halos, and with DESI ELGs \citep{Rocher2023b}.

\section{Cosmological fit} \label{sec:cosmofit}
In the current analysis, perturbative and semi-analytical models that are aimed at extracting cosmological information from the observations were tested intensively against galaxy mock catalogues. In eBOSS analysis \citep{Rossi2021,Alam2021}, different HOD models were used in order to quantify the modelling systematics of redshift-space distortion models. The authors found consistent results amongst a large variety of HOD models that validated the accuracy of the theoretical models.
\subsection{Theoretical models}
In the analyses of \cite{delatorre2017,Jullo2019}, the measured anisotropic clustering $\xi(r_{\perp},r_{\parallel})$  at an effective redshift $z$ was combined at the likelihood level with lensing measurements of $\Delta \Sigma$ in order to provide direct constraints on the growth rate of the structure $f(z)$, the amplitude of matter fluctuation $\sigma_8(z)$, and the present time matter density $\Omega_{\mathrm{m}}$. These studies extracted information at scales below $3 \, h^{-1}$Mpc, where our HOD models, as illustrated in Fig. \ref{fig:main_LRG_plot}, show significant differences that might introduce biases when quantifying modelling systematics of $\Delta \Sigma$. 

 In this study, We performed the cosmological analysis of galaxy catalogues populated with the standard and extended HOD framework in order to test the accuracy of a combined analysis of galaxy clustering and galaxy-galaxy lensing. We considered as a redshift space model the TNS \citep{taruya10} model extended to non-linearly biased tracers, hereafter referred to as TNS, described and robustly tested in detail in \cite{bautista21}. The expression for the redshift space power spectrum of biased tracers is given by 
\begin{multline}
P^s(k,\nu) = D(k\nu\sigma_v) \big[ P_{\rm gg}(k) + 2\nu^2fP_{\rm{g} \theta}(k) + \nu^4f^2 P_{\theta\theta}(k) + \\
C_A(k,\nu,f,b_1) + C_B(k,\nu,f,b_1) \big],
\label{eq:psg}
\end{multline}
where $k$ is the wave-vector norm, $\nu$ is the cosine angle between the wave-vector and the l.o.s, $\theta$ is the divergence of the velocity field $\mathbf{v}$ defined as 
$\theta = -\nabla {\bf \cdot v}/(aHf)$, and $f$ is the linear growth rate parameter. The terms $C_A(k,\nu,f)$ and $C_B(k,\nu,f)$ are the two correction terms given in \citet{taruya10}, which reduce to one-dimensional integrals of the linear matter power spectrum. The damping function $D(k\nu\sigma_v)$ describes the Fingers-of-God effect induced by random motions in virialised systems, inducing a damping effect on the power spectra. We adopted a Lorentzian form, $D(k,\nu,\sigma_v) = (1+k^2\nu^2\sigma_v^2/2)^{-2}$, where $\sigma_v$ represents an effective pairwise velocity dispersion treated as an extra nuisance parameter.
The expressions of the real space quantities $P_{\mathrm{gg}}, P_{\mathrm{g}\theta}$, and $P_{\mathrm{gm}}$ are given following the bias expansion formalism of \cite{assassi14}:
\begin{equation}
\begin{aligned}
P_{\mathrm{gg}}(k)= & b_1^2 P_{\delta \delta}(k)+b_2 b_1 I_{\delta^2}(k)+2 b_1 b_{\mathcal{G}_2} I_{\mathcal{G}_2}(k) \\
& +2\left(b_1 b_{\mathcal{G}_2}+\frac{2}{5} b_1 b_{\Gamma_3}\right) F_{\mathcal{G}_2}(k)+\frac{1}{4} b_2^2 I_{\delta^2 \delta^2}(k) \\
& +b_{\mathcal{G}_2}^2 I_{\mathcal{G}_2 \mathcal{G}_2}(k) \frac{1}{2} b_2 b_{\mathcal{G}_2} I_{\delta_2 \mathcal{G}_2}(k), \\
P_{\mathrm{g} \theta}(k)= & b_1 P_{\delta \theta}(k)+\frac{b_2}{2} I_{\delta^2 \theta}(k)+b_{\mathcal{G}_2} I_{\mathcal{G}_2 \theta}(k) \\
& +\left(b_{\mathcal{G}_2}+\frac{2}{5} b_{\Gamma_3}\right) F_{\mathcal{G}_2 \theta}(k) , \\
P_{\mathrm{gm}}(k)= & b_1 P_{\delta \delta}(k)+\frac{b_2}{4} I_{\delta^2}(k)+b_{\mathcal{G}_2} I_{\mathcal{G}_2}(k) \\
& +\left(b_{\mathcal{G}_2}+\frac{2}{5} b_{\Gamma_3}\right) F_{\mathcal{G}_2}(k),
\end{aligned}
\end{equation}
where each function of $k$ is a one-loop integral of the linear matter power spectrum $P_{\mathrm{lin}}$, whose expression can be found in \cite{Simonvic2018}. The expressions of $I_{\delta^2 \theta}(k)$, $I_{\mathcal{G}_2 \theta}(k)$, and $F_{\mathcal{G}_2 \theta}(k)$ are the same as $I_{\delta^2}(k)$, $I_{\mathcal{G}_2 }(k)$, and $F_{\mathcal{G}_2}(k)$ except that the $G_2$ kernel replaces the $F_2$ kernel in the integrals. To estimate the velocity divergence power spectra $P_{\theta \theta}$ and $P_{\delta \theta}$, we used the \cite{bel19} fitting function, which depends solely on $\sigma_8$:
\begin{equation}
\begin{aligned}
& P_{\theta \theta}(k)=P_{\rm lin}(k) e^{-k\left(a_1+a_2 k+a_3 k^2\right)}, \\
& P_{\delta \theta}(k)=\left(P_{\delta \delta}(k) P_{\text {lin }}(k)\right)^{\frac{1}{2}} e^{-\frac{k}{k_\delta}-b k^6},
\end{aligned}
\end{equation}
with the expressions of each parameter given in  \cite{bel19,bautista21}.
The linear matter power spectrum was estimated with \texttt{camb} \citep{camb2011},\footnote{wrapped around cosmoprimo \\ \url{https://cosmoprimo.readthedocs.io/en/latest/index.html}} while the non-linear power spectrum $P_{\delta \delta}$ was estimated with the latest version of \texttt{HMcode} \citep{Mead2021}. This is different from \cite{bautista21,PAVIOTBB2022}, where we used the response function formalism \texttt{respresso} described in \cite{nishimichi17}. Indeed, the latest version of \texttt{HMcode} provides nearly indistinguishable (at least in configuration space) baryon acoustic oscillation damping due to non-linear effects while also matching the power spectrum of \texttt{respresso} at a higher $k$. 

The TNS correlation function multipole moments were finally determined by performing the Hankel transform on the model power spectrum multipole moments,
\begin{equation} \label{eq:multipoleTNS}
\xi^\mathrm{TNS}_\ell(s) = i^\ell \frac{2\ell+1}{2} \int \der k \frac{k^2}{2 \pi^2} j_\ell(ks) \int_{-1}^{1} \der \nu\, P^s(k, \nu) L_\ell(\nu),
\end{equation}
where $j_\ell$ and  $L_\ell$ denote the spherical Bessel function and Legendre polynomials of order $\ell$. The Hankel transform, that is, the outer integral in the above equation, was performed rapidly using the FFTlog algorithm \citep{hamilton00}. Under the limber approximation \cite{limber53}, the expression of the galaxy-galaxy lensing signal is given by \cite{Marian2015}
\begin{equation}
\Delta \Sigma_{\mathrm{gm}} (r_{\mathrm{p}})= \rho_{\mathrm{m}} \int \frac{\der^2 k_{\perp}}{(2 \pi)^2} P_{\mathrm{gm}}\left(k_{\perp}, z_l\right) J_2\left(k_{\perp} R\right),
\end{equation}
with $J_2$ as the second-order Bessel function of first kind and $\rho_{\mathrm{m}}$ as the co-moving matter density, constant over cosmic time. In order to remove the contribution of non-linear modes, we computed the annular differential excess surface density (ASAD) estimator given by \citep{Baldauf2010}
\begin{equation} \label{eq:Upsilondef}
\Upsilon\left(r_{\mathrm{p}}, R_0\right)=\Delta \Sigma_{\mathrm{gm}}\left(r_{\mathrm{p}}\right)-\left(\frac{R_0}{r_{\mathrm{p}}}\right)^2 \Delta \Sigma_{\mathrm{gm}}\left(R_0\right),
\end{equation}
where $R_0$ corresponds to a cut-off radius, below which the signal is damped. 

In addition, we parametrised the Alcock-Paczy\'nski (AP) distortions \citep[][]{alcock79} induced by the assumed fiducial cosmology in the measurements via two dilation parameters that scale transverse, $\alpha_{\perp}$, and radial, $\alpha_{\parallel}$, separations. These quantities are related
to the co-moving angular diameter distance, $D_M =
(1+z)D_A(z)$, and Hubble distance, $D_H = c/H(z)$, respectively, as
\begin{align}
\alpha_{\perp} &= \frac{D_M(z_{\rm eff})}{D_M^{\rm fid}(z_{\rm eff})},
\label{eq:aperp} \\
\alpha_{\parallel} &=  \frac{D_H(z_{\rm eff})}{D_H^{\rm fid}(z_{\rm eff})},
\label{eq:apara}
\end{align}
where $c$ is the speed of light in the vacuum and $z_{\rm eff}$ is the effective redshift of the sample. We applied these dilation parameters to the theoretical TNS power spectrum $P^{s}(k,\nu)$ in Eq. \eqref{eq:multipoleTNS} so that $P^{s}(k,\nu) \rightarrow P^{s}(k',\nu')$, where
\begin{eqnarray}
\label{eq:AP1}k' &=& \frac{k}{\alpha_\perp}\left[1+\nu^2\left(\frac{1}{F_{\rm AP}^2}-1\right)\right]^{1/2},   \\
\label{eq:AP2}\nu'\ & = & \frac{\nu}{F_{\rm AP}}\left[ 1+\nu^2\left(\frac{1}{F_{\rm AP}^2}-1\right)\right]^{-1/2},
\end{eqnarray}
and $F_{\rm AP}=\alpha_\parallel/\alpha_\perp$.
The terms $\Delta \Sigma$ and $\Upsilon$, being projected quantities, are only distorted by the perpendicular distortion factor 
\begin{equation}
\begin{aligned}
\Delta \Sigma (R') &= \Delta \Sigma (\alpha_{\perp} R), \\
\Upsilon (R') &= \Upsilon (\alpha_{\perp} R). \\
\end{aligned}
\end{equation}
Given a theoretical model for $\xi_{\ell}$ and $\Delta \Sigma$ ($\Upsilon$), we could perform a cosmological analysis assuming a Gaussian likelihood as defined in Eq. \eqref{eq:likelihood}. Here, our data vector is the concatenation of the redshift-space multipoles with the galaxy-galaxy lensing signal, with the free parameters of the model $\theta \subset (b_1,b_2,b_{\Gamma_3},f,\sigma_8,\alpha_{\perp},\alpha_{\parallel},\sigma_v)$, with $b_{\mathcal{G}_2}$ being fixed to the Lagrangian prescription as $b_{\mathcal{G}_2} = -2/7(b_1 - 1)$ \citep{Saito2014}. It is worth noting that as the TNS corrections and the galaxy bias expansion are functions of the linear power spectrum only, these can be rescaled at each iteration by the normalisation $\sigma_8$. The non-linear power spectrum evaluated with \texttt{HMcode} was first sampled into a broad range of $\sigma_8$ and then interpolated at any value of $\sigma_8$ within our prior. Specifically, we computed $P_{\delta \delta}$ in the $\sigma_8$ range [0.1,1.2] with a step size of $\der \sigma_8 = 0.025$. Hence, the only quantities that we re-computed at each iteration were the real space quantities $P_{\theta \theta}$ and $P_{\delta \theta}$. We assumed as a fiducial cosmology, the cosmology of the {\sc uchuu} simulation such that AP distortions should be equal to unity. 
\subsection{Covariance matrices}
The covariance matrices were estimated with analytical prescriptions. The Gaussian covariance for the multipoles of the correlation function is given by \citep{Grieb2016}
\begin{equation} \label{eq:gaussiancovxil}
C_{\ell_1 \ell_2}^{\xi}\left(s_i, s_j\right)=\frac{i^{\ell_1+\ell_2}}{2 \pi^2} \int_0^{\infty} k^2 \sigma_{\ell_1 \ell_2}^2(k) \bar{\jmath}_{\ell_1}\left(k s_i\right) \bar{\jmath}_{\ell_2}\left(k s_j\right) \mathrm{d} k,
\end{equation}
where the $\bar{\jmath}_{\ell}$ correspond to bin-averaged spherical bessel functions,
\begin{equation} \label{eq:sphericalbesselavg}
\bar{\jmath}_{\ell}\left(k s_i\right) \equiv \frac{4 \pi}{V_{s_i}} \int_{s_i-\Delta s / 2}^{s_i+\Delta s / 2} s^2 j_{\ell}(k s) \mathrm{d} s,
\end{equation}
with $V_{si}$ as the volume element of the shell.
The expression for $\sigma_{\ell_1 \ell_2}^2(k)$ is given by 
\begin{equation}
\begin{aligned}
\sigma_{\ell_1 \ell_2}^2(k) \equiv & \frac{\left(2 \ell_1+1\right)\left(2 \ell_2+1\right)}{V_{\mathrm{s}}} \\
& \times \int_{-1}^1\left[P(k, \mu)+\frac{1}{\bar{n}}\right]^2 \mathcal{L}_{\ell_1}(\mu) \mathcal{L}_{\ell_2}(\mu) \mathrm{d} \mu,
\end{aligned}
\end{equation}
with $V_s$ as the volume of the periodic box.
The expression for the Gaussian covariance of $\Delta \Sigma$ is given by \cite{Marian2015}
\begin{equation} \label{eq:gaussiancovds}
\begin{aligned}
& \operatorname{Cov}[\widehat{\Delta \Sigma}]\left(R_i, R_j\right)=\frac{1}{A_s}\left(\rho_{\mathrm{m}}\right)^2 \int \frac{\der^2 k_{\perp}}{(2 \pi)^2} \bar{J}_2\left(k_{\perp} R_i\right) \bar{J}_2\left(k_{\perp} R_j\right) \\
& \times\left\{\left[P_{\delta \delta}\left(k_{\perp}\right)+\frac{1}{\bar{n}_p}\right]\left[P_{\mathrm{gg}}\left(k_{\perp}\right)+\frac{1}{\bar{n}_g}\right]+P_{\mathrm{gm}}^2\left(k_{\perp}\right)\right\},
\end{aligned}
\end{equation}
\begin{table*}
\caption{Result of the cosmological fit from the optimal configurations determined from Fig. \ref{fig:TNSfitDS} and from Fig. \ref{fig:TNSfitUp} for the extended HOD model.}
\begin{tabular}{l|cccc}
\hline
\hline
Parameter  &  $\Delta \alpha_{\perp}$ & $\Delta \alpha_{\parallel}$ & $\Delta f$ & $\Delta \sigma_8$ \\
\hline
\noalign{\vskip 1pt}
LRG $\xi_{\ell}$ $\sigma_8$ fixed & $-0.012 \pm 0.010$ & $0.006 \pm 0.018$  & $-0.035 \pm 0.039$ & \\
LRG $\xi_{\ell}$ $\sigma_8$ free & $-0.013 \pm 0.010$ & $0.003 \pm 0.018$  & $0.325 \pm 0.277$ & $-0.155 \pm 0.099 $\\
LRG $\xi_{\ell}$ + $\Delta \Sigma$  $r_{\mathrm{min}} = 3h^{-1}$Mpc & $-0.013 \pm 0.009$ & $0.004 \pm 0.017$ & $-0.048 \pm 0.051$ & $0.007 \pm 0.026$ \\
LRG $\xi_{\ell}$ + $\Upsilon$   $r_{0} = 1.2 h^{-1}$Mpc & $-0.014 \pm 0.009$ & $-0.002 \pm 0.017$ & $-0.014 \pm 0.054$ & $-0.003 \pm 0.028$ \\
ELG $\xi_{\ell}$ $\sigma_8$ fixed & $-0.004 \pm 0.010$ & $0.002 \pm 0.016$ & $-0.012 \pm 0.026$ & \\
ELG $\xi_{\ell}$ $\sigma_8$ free &  $-0.007 \pm 0.010$ & $0.008 \pm 0.017$ & $0.341 \pm 0.257$ & $-0.128 \pm 0.077 $\\
ELG $\xi_{\ell}$ + $\Delta \Sigma$   $r_{\mathrm{min}} = 1 h^{-1}$Mpc & $-0.007 \pm 0.010$ & $0.001 \pm 0.015$ & $0.062 \pm 0.057$ &  $-0.033 \pm 0.025$ \\
ELG $\xi_{\ell}$ + $\Upsilon$  $r_{0} =  1 h^{-1}$Mpc & $-0.008 \pm 0.010$ & $0.001 \pm 0.016$ & $0.072 \pm 0.062$ & $-0.035 \pm 0.027$ \\
\end{tabular}
\label{tab:result_TNS_fit}
\end{table*}  
where $A_s$ is the surface area of the box. The $\bar{J}_2$ corresponds to bin-averaged Bessel functions, computed in the same way as in Eq. \eqref{eq:sphericalbesselavg} by integrating over the surface element instead. In Eqs \eqref{eq:gaussiancovxil}-\eqref{eq:gaussiancovds}, we assumed linear theory for the power spectra such that $P_{\delta \delta} = P_{\mathrm{lin}}$, $P_{\mathrm{gm}} = b_1 P_{\delta \delta}$, $P_{\mathrm{gg}} = b_1^2 P_{\delta \delta}$, and $P(k,\mu) = b_1^2 P_{\mathrm{gg}}(k)\left(1+\beta^2 \mu^2\right)^2$ \citep{Kaiser1987}, with $\beta = f/b$. We inferred the value of $f$ from the {\sc uchuu} cosmology at the effective redshift of the LRG and ELG sample, respectively $z = 0.7$ and $z=0.86$, while the values of $b_1$ were taken from \cite{bautista21,tamone20}. The shot-noise terms $\bar{n}_p$ and $\bar{n}_g$ correspond to the number density of dark matter particles and galaxies in the simulation box. No analytical cross-covariance exists between $\Delta \Sigma$ and the $\xi_{\ell}$, so we approximated it to be zero. To estimate the covariance of $\Upsilon$, we generated a sample of a thousand $\Delta \Sigma$ from a multivariate normal distribution, given the $\Delta \Sigma$ covariance estimate. We then applied Eq. \eqref{eq:Upsilondef} to each fake sample by considering its mean value for $\Delta \Sigma (R_0)$. This had the effect of reducing the noise in the final covariance estimation.  
\begin{figure*} 
    \centering    
    \sidecaption
    \includegraphics[width=12cm]{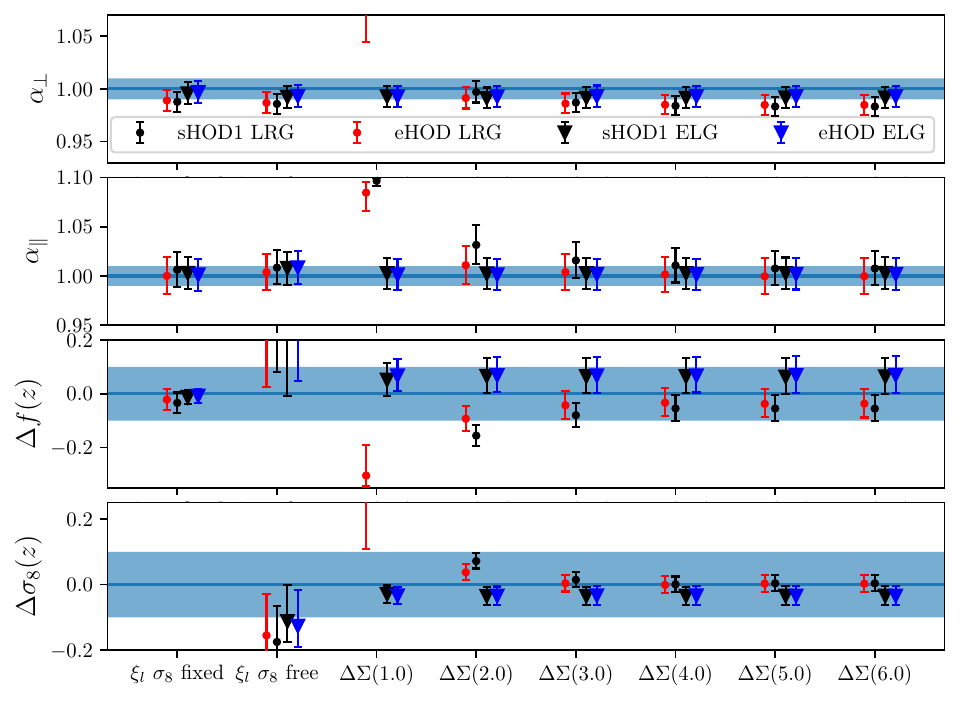}
    \caption{Results from the cosmological fit. Circle points with error bars correspond to the fit of the LRG sample at redshift $z=0.70$ populated with the standard and extended HOD framework, while the triangles correspond to the fit of ELGs at redshift $z=0.86$. The value of $f$ and $\sigma_8$ for LRGs (ELGs) are 0.814 (0.849) and 0.570 (0.529) at the redshift $z=0.70 (0.86)$. Values in parentheses correspond to the minimum scale used in the fit in megaparsec per h. The shaded area corresponds to a $10 \%$ variation around the fiducial value of each parameter.
      }
\label{fig:TNSfitDS}
\end{figure*}
\begin{figure*}
    \centering    
    \sidecaption
    \includegraphics[width=12cm]{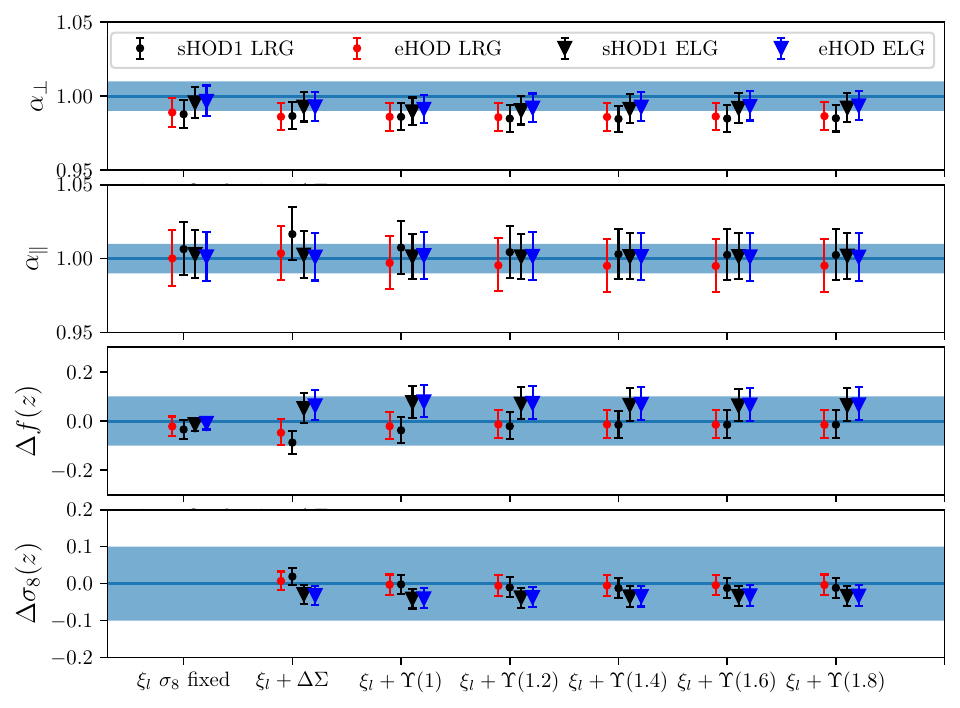}
    \caption{Same as Fig. \ref{fig:TNSfitDS} but this time considering the annular excess surface density $\Upsilon$ instead. 
      } 
 \label{fig:TNSfitUp}
\end{figure*}
\subsection{Analysis choices}
We measured the multipoles of the correlation function, the monopole, quadrupole, and hexadecapole with bins of size 5 $h^{-1}$ Mpc in the range between 0 and 150 $h^{-1}$ Mpc. We measured $\Delta \Sigma$ with 22 and 16 logarithm bins in the range between $10^{-1}$ and $10^{2}$ $h^{-1}$ Mpc for LRGs and ELGs, respectively. This binning was chosen for each sample given the S/N of the observations. Since our redshift-space model follows the exact same implementation as in \cite{bautista21}, we performed a fit of the multipoles on the scales determined in their analysis, which provides unbiased cosmological constraints for an LRG-like sample. These scale cuts were determined by testing the TNS model against the BOSS N-body {\sc Nseries} mocks, which were produced to match the clustering amplitude of CMASS galaxies. Since the BOSS CMASS and eBOSS LRG samples have a similar bias, the scale cuts determined with the {\sc Nseries} mocks were used to determine the cosmological constraints on the DR16 eBOSS LRG sample \citep{bautista21}. We therefore used the minimum fitting range determined in their analysis, and we fit the multipoles (monopole, quadrupole, and hexadecapole) between 25 and 140 $h^{-1}$ Mpc for the LRG sample. In theory, one could fit ELGs clustering down to smaller scales, as these galaxies are i) at a higher redshift and ii) located in lower dense regions. However, we kept the same range in our fitting procedure between the two samples, with the scale cut for an ELG-like sample being conservative. We tested the accuracy of the TNS model as a function of different minimum scales used when fitting either $\Delta \Sigma$ or $\Upsilon$. The maximum scale that we used is 80 $h^{-1}$ Mpc, as Limber approximation breaks down on larger scales. The fits were performed on the average of ten HOD realisations to be robust to stochastic variation due to halo occupation.
\subsection{Results}
We present in Fig. \ref{fig:TNSfitDS} the results of the MCMC chains of the combined $\xi_{\ell}  + \Delta \Sigma$ fit. Here, the inference pipeline was done with the \texttt{nautilus} sampler\footnote{\url{https://nautilus-sampler.readthedocs.io/en/latest/}} \citep{nautilus}. Each chain was stopped after reaching an effective number of 200,000 points, and sampled points in the exploration phase were discarded. We present in Table \ref{tab:result_TNS_fit} the corresponding constraints. As expected, we obtained unbiased constraints when fitting only the clustering while fixing the value of $\sigma_8(z)$ to the fiducial value of the simulation. However, in the scenario where $\sigma_8$ was allowed to vary, we observed a strong degeneracy between $f$ and $\sigma_8$ that is not captured properly by the model given the amount of information present in the multipoles. It is only when combining anisotropic clustering with galaxy-galaxy lensing measurements that one can break this degeneracy. 
\begin{figure}
    \centering    
    \includegraphics[width=1\columnwidth]{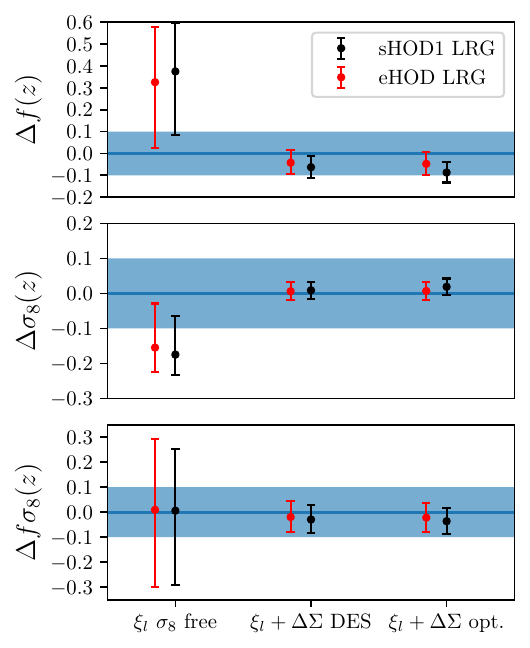}
    \caption{Comparison between a fit to the multipoles $\xi_{\ell}$ only (left) and a fit to the combined $\xi_{\ell} + \Delta \Sigma$ data vector. We present constraints for the scale cuts used for $\gamma_t$ in the DES y3 analysis \citep{Krause2021}, $r_{\mathrm{min}}$ = 6 $h^{-1}$Mpc, and for the optimal configuration determined in this work, $r_{\mathrm{min}}$ = 3 $h^{-1}$Mpc.
      } 
 \label{fig:TNSfs8}
\end{figure}
For the LRG mocks, our model provides unbiased constraints on each cosmological parameter, regardless of the HOD, on scales above $3 h^{-1}$ Mpc for $\Delta \Sigma$, with the constraints on the eHOD that seem to converge faster towards consistent results. This is surprising given the observed difference in Fig. \ref{fig:main_LRG_plot}. The main difference comes from the goodness of fit, with the TNS model providing a better description of the $\Delta \Sigma$ measured with the eHOD galaxy population, $\Delta \chi^2 \approx 20$ compared to the sHOD. The fact that the TNS model provides a good fit, $\chi^2_r = 0.8$, for the eHOD measurements suggests that our model is also capable of yielding a good description of the observations down to a scale of $\approx$ 3 $h^{-1}$Mpc, in agreement with the result of \cite{delatorre2017}.

It is important to note that these results provide a quantitative description of how small-scale variations of $\Delta \Sigma$ impact cosmological analysis. Indeed, we did not consider satellite segregation and baryonic feedback in this work. Including these effects would most likely impact the evolution of cosmological constraints as a function of scale but would not change the observed trends, for example, a better description of an eHOD $\Delta \Sigma$ with perturbative models.

For ELGs, one can see that both HOD models provide constraints indistinguishable from one another, as both methods provide very similar clustering and lensing measurements. Surprisingly, we find that the model provides unbiased and consistent constraints over all the considered range down to $r_{\mathrm{min}}$ = 1$h^{-1}$Mpc. This means that our estimator of $P_{\mathrm{gm}}$, which is built upon a one-loop biased expansion of the non-linear matter power spectrum estimated with \texttt{HMcode}, is valid up to non-linear scales to describe ELGs, at least at an effective redshift of $z=0.86$. This might be related to their spatial distribution. Indeed, \cite{GonzalezPerez2020} found that half of DESI ELGs reside in filaments and a third in sheets, where density fluctuations are small compared to the nodes where LRGs usually lie. This is consistent with the results of \cite{Hadzhiyska2021}, who found that ELGs are two times as likely to reside in sheets compared to a mass-selected sample. In addition, ELGs occupy lower mass halos, $\log M \approx 12.2 \, h^{-1}M_{\odot}$, compared to LRGs, with $\log M$  in the range $12.8-13.2 \, h^{-1}M_{\odot}$, such that the one-halo term contribution to $\Delta \Sigma$ is expected to be smaller for ELGs at around $1 \, h^{-1}$Mpc. This explains why our bias perturbative expansion provides a good description of the galaxy-galaxy lensing signal of the ELGs down to small scales. As a baseline configuration, we adopted $r_{\mathrm{min}}=1$ and $3 \, h^{-1}$Mpc for ELGs and LRGs, respectively.

Our analysis demonstrates the constraining power of $\Delta \Sigma$ in combination with anisotropic clustering measurements, as illustrated in Fig. \ref{fig:TNSfs8}. This methodology provides unbiased constraints on the cosmological parameters $\alpha_{\perp},\alpha_{\parallel}, f$, and $\sigma_8$, improving the precision compared to a $\xi_{\ell}$ fit only by a factor of about six and three for $f$ and $\sigma_8$, respectively. However, we emphasize that this conclusion is obviously optimistic, as the variance in observation is dominated by the shape-noise contribution \citep{Gatti2021}, which is not considered in this work. We leave to future work a more detailed analysis with a realistic light-cone. We compared the optimal LRG configuration determined in this work, $r_{\mathrm{min}} = 3 \, h^{-1}$Mpc, to the one determined in DES analysis \citep{Krause2021},  $r_{\mathrm{min}} = 6 \, h^{-1}$Mpc. We observed similar constraints for the two configurations, with a small $3\%$ improvement in precision for $\sigma_8$. When moving down to highly non-linear scales, the two-point function captures less and less information such that the gain in precision becomes smaller.

We present in Fig. \ref{fig:TNSfitUp} the results of the fits when combining the $\xi_{\ell}$ with $\Upsilon$ compared to the baseline configuration determined from Fig. \ref{fig:TNSfitDS}. Since the ELG fits for $\Delta \Sigma$ provide consistent results up to $1 \, h^{-1}$Mpc, the constraints when fitting $\Upsilon$ should be the same, regardless of the value of $R_0$, which is exactly what we observed in this figure. The trend is similar for LRGs, with consistent results when fitting $\Upsilon$ with a value of $R_0$ above $1.2 \, h^{-1}$Mpc. As a baseline configuration, we therefore adopted $R_0 = 1$ and $1.2 \, h^{-1}$Mpc for ELGs and LRGs, respectively. We present in Table \ref{tab:result_TNS_fit} the corresponding constraints with our baseline configuration for $\Delta \Sigma$ and $\Upsilon$. Both methods provide similar constraints with consistent results at the 1$\sigma$ level. The difference in $\chi^2$ is smaller between the extended and standard framework, $\Delta \chi^2 = 5$, as observed differences in $\Delta \Sigma$ are smoothed out with the annular excess surface density estimator. It seems that some information is lost when using $\Upsilon$ as a proxy for galaxy-galaxy lensing since the constraints on $f$ and $\sigma_8$ are larger by 5 to 8$\%$. However, this effect is marginal, below $1\%$, compared to the statistical precision. We therefore conclude that both statistics, either $\Delta \Sigma$ or $\Upsilon$, can be used interchangeably when doing a joint cosmological analysis of galaxy clustering and lensing.  

\section{Conclusion}
\label{sec:conclusion}
In this paper, we have revisited the extended HOD framework introduced in \cite{Yuan2022}, built on top of the state-of-the-art {\sc uchuu} simulation. This HOD model includes internal (halo concentration and shape) and environmental (local density and local shear) dependencies of central and satellite occupation. We applied this model to two eBOSS tracers, ELGs in the range $0.6 < z < 1.1$ and LRGs in the range $0.6 < z < 1.0$, by modelling the projected correlation $w_p$ and the galaxy-galaxy lensing signal $\Delta\Sigma$ from highly non-linear ($\approx$ Mpc scale) up to linear scales.\\ 
We find that the extended HOD model provides a better description of both the clustering and lensing signal of LRGs, reducing the mean observed discrepancy on $\Delta \Sigma$ from 36$\%$ to 12$\%$ compared to the standard HOD model based on halo mass only. Our results demonstrate the necessity of including in the HOD framework environment-based secondary parameters with a strong detection, at the 3$\sigma$ level, of LRGs preferentially occupying anisotropic and denser environments. As internal halo property, we tested the concentration and shape of the dark matter halo, and we find that both properties can be used to improve the HOD model. For ELGs, we find that given our statistical precision, either the standard or extended HOD framework provides consistent and robust modelling for both the clustering and lensing signal, with a slight preference for ELGs living in denser regions, in agreement with previous results \citep{Artale2018,Delgado2022}. 

We then conducted a cosmological analysis of the redshift-space correlation function and the galaxy-galaxy lensing signal for both HOD frameworks using a perturbative model. For LRGs, we find that our model provides unbiased and consistent constraints on the linear growth rate $f$, on the amplitude of fluctuations $\sigma_8$, and on the Alcock-Paczynski parameters on scales above $3 \, h^{-1}$Mpc and $1.2 \, h^{-1}$Mpc when fitting the galaxy-galaxy lensing signal $\Delta \Sigma$ and the annular excess surface density $\Upsilon$, respectively. In addition, we find that our model provides a better description of the extended HOD framework, suggesting that the observed eBOSS galaxy-galaxy lensing signal can be accurately modelled down to scales below $5 \, h^{-1}$Mpc, where significant discrepancies can be observed between the HODs and the data. Moreover, we find for ELGs that we can derive robust constraints from the galaxy-galaxy lensing signal down to very small scales, $r_{\mathrm{min}} = 1.0 \, h^{-1}$Mpc, regardless of the estimator of the galaxy-galaxy lensing signal. This analysis will serve as a cornerstone for a companion paper that will extract cosmological information from the eBOSS galaxy-galaxy lensing signal and pave the way for future cosmological analysis of joint clustering and galaxy-galaxy lensing measurement for future cosmological experiments such as DESI, \textit{Euclid} \citep{EUCLID2024}, and the Vera C. Rubin Observatory.

\begin{acknowledgements}
The authors thank the referee for the useful comments that considerably improved the quality of this work. RP is supported by a DIM-ACAV+ fellowship from {\sl Region Ile-de-France}. SC acknowledges financial support from Fondation MERAC and by the SPHERES grant ANR-18-CE31-0009 of the French {\sl Agence Nationale de la Recherche}. In addition to the packages already quoted in the main text, we also acknowledge the use of \texttt{numpy} \citep{numpy} and \texttt{scipy} \citep{2020SciPy-NMeth}. Plots were made with \texttt{matplotlib}  \cite{matplotlib} while corner plots were made with \texttt{pygtc} 
 \cite{pygtc}. 
\end{acknowledgements}

\bibliographystyle{aa}

\bibliography{ref} 
\begin{appendix}
\section{Robustness of the analysis} \label{sec:appendixA}
\begin{figure}[h!]
    \centering    
    \includegraphics[width=\columnwidth]{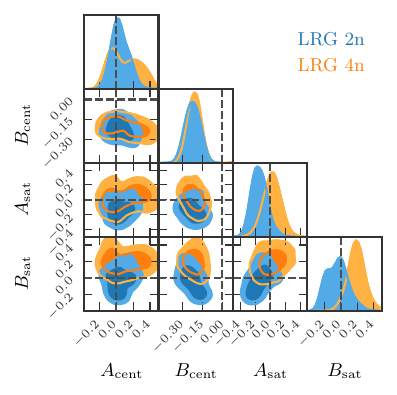}
    \caption{Full posterior distributions of the assembly bias and environmental parameters in our baseline configuration. We also present our result with a different number density from the one defined in the main text. The dependence of halo occupation on the environment is unchanged.
      }
    \label{fig:contours_AB_4n}  
\end{figure}

In our main analysis, we fix the number density in the simulation box to be twice the mean number density of LRG, where $\bar{n}_{\mathrm{gal }}^{\mathrm{LRG}} = 7.8 \times 10^{-5} h^{3} \mathrm{Mpc}^{-3}$. This choice was motivated by the fact that increasing the number density provides higher S/N measurements such that averaging over ten realisations is enough to get robust likelihood estimates. This is particularly important for numerical speed, as we do not use any emulator to compute $\Delta \Sigma$. In Fig. \ref{fig:contours_AB_4n}, we present the posterior distributions, this time fixing the number density of galaxy in the box to 4 times the number density of LRG. As we can see, the environmental dependence on halo occupation is unchanged.

\begin{figure}[h!]
    \centering    
    \includegraphics[width=\columnwidth]{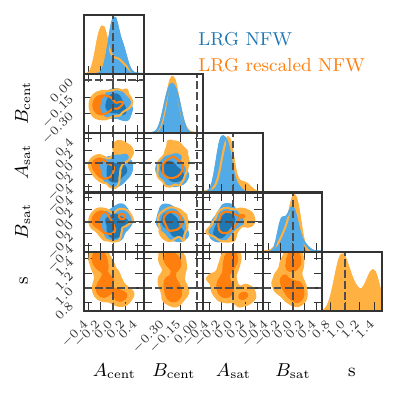}
    \caption{Full posterior distributions of the assembly bias and environmental parameters in our baseline configuration. We also present our result with a modified NFW profile. The dependence of halo occupation on the environment is unchanged.
      }
    \label{fig:contours_AB_withs}  
\end{figure}

Additionally, we modified the NFW radial profile as done in \cite{Avila2020}. We introduce an additional nuisance parameter $s$, and we modify the concentration of each halo by $c_{\textrm{new}} = s \times c$. We allowed $s$ to vary in the range $[0.5,1.5]$. We present the result in Fig. \ref{fig:contours_AB_withs}. This additional degree of freedom does not introduce significant change in the environmental dependence of halo occupation. Additionally, we can see that the data seems to favour an NFW profile.
\FloatBarrier
\FloatBarrier
\section{Additional parameters}\label{sec:appendixB}
\begin{figure}[h!]
    \centering 
    \includegraphics[width=1\columnwidth]{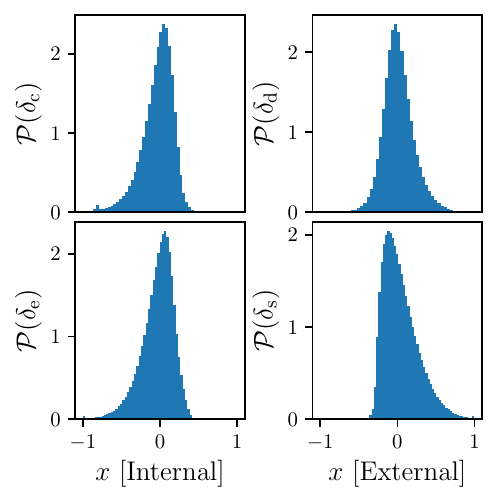}
    \caption{Distribution of internal and external halo properties as defined in Sect. \ref{sec:ABpara}. The left panels present the distribution of concentration (c) and ellipticity (e), while the right panels present the local density (d) and local shear (s).}
    \label{fig:distributionAB}
\end{figure}

\begin{figure}[h!]
    \centering 
    \includegraphics[width=1\columnwidth]{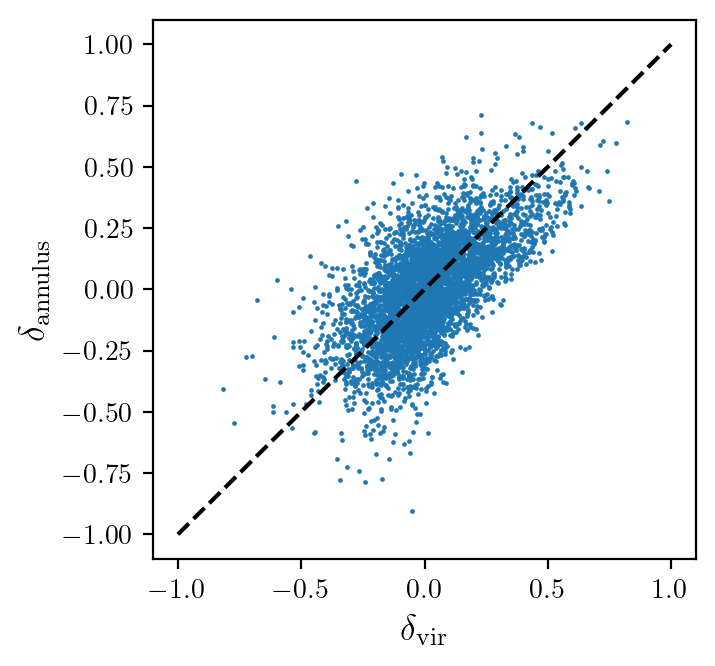}
    \caption{Comparison between the local density estimated on a mesh with a Gaussian filter of size $R$ = 2.25 $r_{\mathrm{vir}}$ and the local density estimated by summing halo and sub-halo masses around $r_{\mathrm{vir}}$ and 5 $h^{-1}$Mpc. The dashed line corresponds to $\delta_{\rm vir} = \delta_{\rm annulus}$.}
    \label{fig:distributiondensityy}
\end{figure}

We present in Fig. \ref{fig:distributionAB} the normalised distribution in the range [-1,1] of internal (concentration and ellipticity) and external (local density and local shear) halo properties on the left and right panel respectively. We present in Fig. \ref{fig:distributiondensityy} the relation between the two different local density estimators described in this work.    
\FloatBarrier
\FloatBarrier
\section{Posterior of the luminous red galaxies halo occupation distribution}\label{sec:appendixC}
\begin{figure*}[h!]
    \centering    
    \includegraphics[width=1.8\columnwidth]{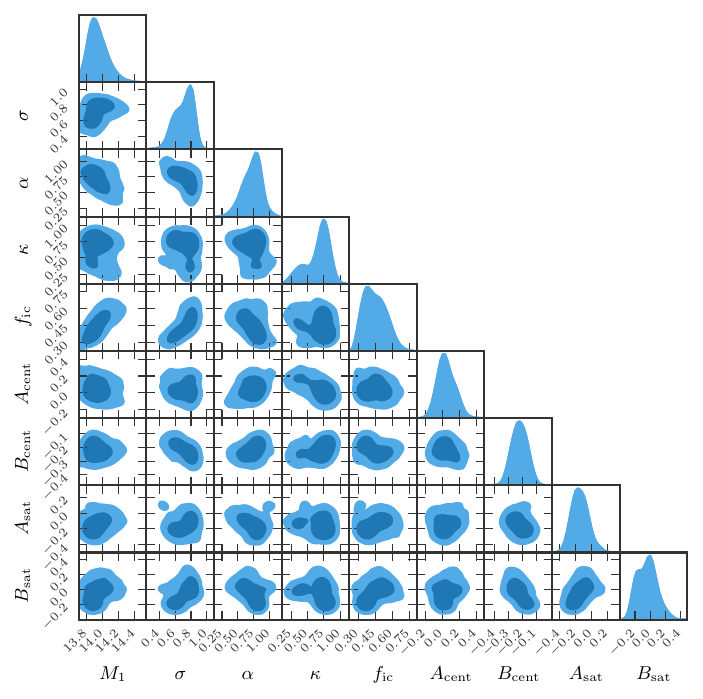}
    \caption{Full posterior distribution of the LRG-extended HOD, including the assembly bias parameters in the fiducial configuration.}
    \label{fig:contours_LRG_FULLBB}  
\end{figure*}

We present here the full posterior contours in Fig. \ref{fig:contours_LRG_FULLBB} in the fiducial configuration. Additionally, we present in Table \ref{tab:result_HOD_LRG_full} the posterior distributions for all configurations presented in Sect. \ref{sec:diffHOD}. We only present $M_{\mathrm{min}}$ for the posterior distributions presented in Fig. \ref{fig:contours_LRG} to save computational time, as $M_{\mathrm{min}}$ is not a free parameter and is computed for each point of the chain given Eq. \eqref{eq:ngal}. In addition, for completeness, we also present the full posterior contours.
\FloatBarrier
\begin{table*}[h!]
  \caption{Result of the Gaussian process emulating the likelihood function after 4000 iterations for the eBOSS LRG sample.}
  \begin{tabular}{lp{1.5cm}p{1.5cm}p{1.5cm}p{1.5cm}p{1.5cm}p{1.5cm}p{1.5cm}p{1.5cm}}
\hline
\hline
Method  &  sHOD1 & sHOD2 & eHOD $\delta_{\mathrm{vir}}$,c & eHOD $T^{ij}_{\mathrm{vir}}$,c & eHOD $T^{ij}_{\mathrm{vir}}$,b/a &  eHOD $\delta_{\mathrm{annulus}}$ ,c & eHOD $\delta_{\mathrm{annulus}}$,b/a & eHOD $T^{ij}_{5\mathrm{Mpc}}$ ,c \\ 
\hline
\noalign{\vskip 1pt}
   $M_{1}$ [13.7,15.0]& $14.17^{+0.13}_{-0.12}$ & $14.02^{+0.07}_{-0.06}$& $14.02^{+0.12}_{-0.09}$ & $13.91^{+0.15}_{-0.14}$  & $13.86^{+0.10}_{-0.09}$  & $13.92^{+0.14}_{-0.11}$ & $13.91^{+0.12}_{-0.11}$  &  $13.94^{+0.13}_{-0.11}$ \\
   $\sigma$ [0.1,1.2] &  $0.53^{+0.11}_{-0.12}$ & $0.40^{+0.11}_{-0.13}$ & $0.50^{+0.11}_{-0.13}$& $0.77^{+0.10}_{-0.12}$ & $0.65^{+0.13}_{-0.13}$ & $0.74^{+0.09}_{-0.14}$  & $0.79^{+0.11}_{-0.10}$ &   $0.79^{+0.10}_{-0.09}$ \\
   $\alpha$ [0.1,1.5] & $1.11^{+0.14}_{-0.15}$ & $1.12^{+0.16}_{-0.10}$ & $0.83^{+0.20}_{-0.20}$ & $0.83^{+0.15}_{-0.14}$ & $0.92^{+0.13}_{-0.19}$ & $0.73^{+0.14}_{-0.18}$ & $0.82^{+0.11}_{-0.11}$  & $1.02^{+0.09}_{-0.16}$  \\
   $\kappa$ [0.1,1.5] &$0.59^{+0.30}_{-0.31}$ & $0.72^{+0.21}_{-0.22}$ & $0.79^{+0.13}_{-0.16}$ & $0.29^{+0.29}_{-0.15}$ & $0.73^{+0.20}_{-0.32}$ & $0.69^{+0.14}_{-0.31}$ & $0.58^{+0.26}_{-0.32}$ & $0.45^{+0.16}_{-0.12}$ \\
   $f_{\mathrm{ic}}$ [0.1,1.0] & $0.64^{+0.20}_{-0.15}$ &  $0.45^{+0.10}_{-0.07}$ &  $0.41^{+0.08}_{-0.07}$ & $0.42^{+0.15}_{-0.18}$ &  $0.34^{+0.08}_{-0.07}$ &$0.45^{+0.11}_{-0.10}$ &  $0.43^{+0.10}_{-0.09}$ & $0.45^{+0.09}_{-0.07}$ \\
   $A_{\mathrm{cent}}$ [-0.5,0.5] &&& $-0.01^{+0.15}_{-0.12}$  & $0.24^{+0.23}_{-0.31}$  & $0.01^{+0.19}_{-0.13}$ & $0.04^{+0.14}_{-0.10}$ & $0.12^{+0.24}_{-0.16}$ & $0.26^{+0.15}_{-0.38}$  \\
   $B_{\mathrm{cent}}$ [-0.5,0.5] &&& $-0.24^{+0.09}_{-0.08}$ & $-0.27^{+0.10}_{-0.09}$ & $-0.20^{+0.07}_{-0.06}$ & $-0.23^{+0.06}_{-0.06}$ & $-0.26^{+0.07}_{-0.08}$ & $-0.20^{+0.06}_{-0.06}$  \\
   $A_{\mathrm{sat}}$ [-0.5,0.5] &&& $0.13^{+0.20}_{-0.31}$ & $-0.21^{+0.36}_{-0.18}$ & $-0.02^{+0.24}_{-0.14}$  & $-0.14^{+0.12}_{-0.11}$ & $0.01^{+0.16}_{-0.32}$ &  $-0.10^{+0.12}_{-0.24}$ \\
   $B_{\mathrm{sat}}$ [-0.5,0.5] &&& $-0.08^{+0.11}_{-0.09}$ & $0.20^{+0.14}_{-0.27}$ & $-0.08^{+0.21}_{-0.10}$ & $-0.02^{+0.12}_{-0.15}$ & $0.04^{+0.21}_{-0.22}$ &  $0.04^{+0.08}_{-0.10}$\\
   \hline
\noalign{\vskip 1pt}
   $M_{\mathrm{min}}$ & $13.28^{+0.09}_{-0.09}$ & & & & & $13.27^{+0.14}_{-0.17}$ & & \\
\end{tabular}
\label{tab:result_HOD_LRG_full}
\tablefoot{ We present here the 16, 50, and 84 percentiles of the posterior distributions. The priors are presented in brackets and correspond to uniform priors within the quoted ranges.}
\end{table*}
\FloatBarrier

\end{appendix}
\end{document}